\documentstyle[12pt,epsfig]{article}
\makeatletter
\def\slash#1{{\mathpalette\c@ncel{#1}}} 
\makeatother

\newcommand\beq{\begin{eqnarray}}
\newcommand\eeq{\end{eqnarray}}

\newcommand\la{\langle}
\newcommand\ra{\rangle}

\newcommand{\sslash}[1]{\not{\!#1}}

\def\Pslash{\rlap/{\mkern-1mu P}}

\def\Sslash{\rlap/{\mkern-1mu S}}
\def\pslash{\rlap/{\mkern-1mu p}}

\def\kslash{\slash{\mkern-1mu k}}
\def\nslash{\slash{\mkern-1mu n}}

\def\Sslash{\slash{\mkern-1mu S}}
\def\xhat{\widehat{x}}
\def\zhat{\widehat{z}}

\def\kvec{\vec{k}}

\def\GFt{\widetilde{G}_F}
\def\GDt{\widetilde{G}_D}
\def\gt{\widetilde{g}}

\begin{document}
\begin{flushright}
\end{flushright}
\vspace*{15mm}
\begin{center}
{\Large \bf Twist-3 Formalism for Single 
\\[2mm] 
Transverse Spin Asymmetry Reexamined: 
\\[4mm] 
Semi-Inclusive Deep Inelastic Scattering}
\vspace{1.5cm}\\
 {\sc Hisato Eguchi$^1$, Yuji Koike$^1$, Kazuhiro Tanaka$^2$}
\\[0.4cm]
\vspace*{0.1cm}{\it $^1$ Department of Physics, Niigata University,
Ikarashi, Niigata 950-2181, Japan}\\
\vspace*{0.1cm}{\it $^2$ Department of Physics, 
Juntendo University, Inba-gun, Chiba 270-1695, Japan}
\\[3cm]

{\large \bf Abstract} \end{center}

We study the single spin asymmetry (SSA) for the pion production in 
semi-inclusive deep inelastic scattering, $ep^\uparrow\to e\pi X$,
in the framework of the collinear factorization.  
We derive the complete cross section formula 
associated with the twist-3 quark-gluon correlation functions 
for the transversely polarized nucleon, including all types of pole
(hard-pole, soft-fermion-pole and soft-gluon-pole) contributions 
which produce the strong interaction phase necessary for SSA.  
We prove that the partonic hard part from each pole contribution
satisfies certain constraints from Ward identities for color gauge invariance. 
We demonstrate that the use of these new constraints 
is crucial to reorganize the collinear expansion of the Feynman diagrams
into manifestly gauge-invariant form so as to
obtain the factorization formula for the cross section
in terms of a complete set of the twist-3 distributions without any double counting.  
It also provides a simpler method for the actual calculation.  
We also present a
simple estimate of SSA 
based on our cross section formula, 
using a model for the ``soft-gluon-pole 
function'' that represents the relevant twist-3 quark-gluon correlation,
and compare the magnitude of the terms involving the derivative of the 
soft-gluon-pole function with that of the
``non-derivative'' terms.

\noindent

\newpage
\section{Introduction}

Understanding of the
large single transverse spin asymmetries (SSAs) observed in $pp$ collisions
in 70s and 80s \,\cite{Lambda,E704} 
has been a big challenge for QCD theorists. (See \cite{Review} for a review.) 
Recent data on SSA at higher energies in $pp$ collisions at 
RHIC\,\cite{STAR,PHENIX,BRAHMS}, and 
SSAs in semi-inclusive deep-inelastic scattering (SIDIS)\,\cite{hermes,compass}
have been providing even more exciting opportunities to clarify 
fundamental mechanisms of SSA as well as the relevant hadron structures.  
Time-reversal invariance in QCD implies that the
nonzero SSA is caused by the interference of the amplitudes which have different
phases.  In addition, the nonzero SSA requires a helicity-flip mechanism 
induced by certain mass scale.
In the framework of the factorization formula with conventional 
twist-2 parton distribution functions
and fragmentation functions, the corresponding effects are provided through 
purely perturbative mechanism and is known to give only tiny SSA\,\cite{KPR}.
Therefore, the observed large SSAs require the extension of the factorization 
formula for high energy semi-inclusive reactions
to incorporate novel nonperturbative functions beyond conventional twist-2
distribution/fragmentation functions,
and those new 
functions should represent
the single helicity flip mechanism 
reflecting the dynamics of chiral symmetry breaking.  

In the literature, two QCD mechanisms
have been developed to
explain the large SSAs,
in conjunction with the corresponding novel nonperturbative
functions. One is based on the use of so-called ``T-odd'' distribution
and fragmentation functions which have explicit
intrinsic transverse momentum $\kvec_\perp$ of partons inside hadrons\,\cite{Sivers,Collins93}.
This mechanism describes the 
SSA as a leading twist effect in the region of small transverse momentum $p_T$  
of the produced hadron.
Factorization formula with the use of
$\vec{k}_\perp$-dependent parton distribution functions
and fragmentation functions 
has been extended to SIDIS\,\cite{JMY05}, applying the method used for 
$e^+e^-$ annihilation\,\cite{CS81} and the Drell-Yan process\,\cite{CSS85}, and
the universality property
of those $k_\perp$-dependent functions
have been examined in great details\,
\cite{Collins02,BJY03,BMP03,CM04,BMP04}.  
Phenomenological applications of the ``T-odd'' functions 
have been also performed to interpret the existing data for SSAs\,\cite{Todd}. 

The other mechanism describes the SSA as a twist-3 effect 
in the collinear factorization\,\cite{ET82,QS91}, and is suited for describing
SSA in the large $p_T$ region.  
In this framework, SSA is connected 
to particular quark-gluon correlation functions on the lightcone. 
This formalism has been applied to SSA in 
$pp$ collisions, such as direct photon production $p^\uparrow p\to \gamma X$\,\cite{QS91},
pion production 
$p^\uparrow p\to\pi X$\,\cite{QS99,KK00,Koike03,KQVY06}, 
the hyperon polarization $pp\to\Lambda^\uparrow X$\,\cite{KK01,hyperon},
and SIDIS $ep^\uparrow\to e\pi X$\,\cite{EKT06}.
Although 
the above two mechanisms describe SSA in different
kinematic regions, it has been known that the soft-gluon-pole function
appearing in the twist-3 mechanism is connected to a $\vec{k}_\perp$ moment of a
``T-odd'' function\,\cite{BMP03,BMT98}.     
More recently, 
it has been also shown in refs. \cite{JQVY06, JQVY06DIS} that 
the two mechanisms give the identical 
SSA in the intermediate $p_T$ region for the Drell-Yan process as well as for the SIDIS,
unifying the two mechanisms.

In our previous paper\,\cite{EKT06}, we studied the SSA 
in SIDIS, $ep^\uparrow\to e\pi X$, applying the twist-3 mechanism.   
The purpose of that study was to derive the
cross section formula which is valid in the region of the large transverse
momentum of the pion, i.e. $p_T\sim Q \gg \Lambda_{\rm QCD}$ with $Q$ the
virtuality
of the exchanged virtual photon, 
and to present an estimate of SSA in SIDIS,
in comparison with the similar twist-3 mechanism that 
reproduces the asymmetry
$A_N$ observed for $pp^\uparrow\to\pi X$.  
There we identified two kinds of terms in the twist-3 cross section,
contributing to SSA in SIDIS: 
\beq
({\rm A}) & & G_F(x_1,x_2)\otimes D(z) \otimes \widehat{\sigma}_{\rm A},\nonumber\\
({\rm B}) & & \delta q(x) \otimes \widehat{E}_F(z_1,z_2) \otimes \widehat{\sigma}_{\rm B},
\label{twist3}
\eeq
where $G_F(x_1,x_2)$ and $\widehat{E}_F(z_1,z_2)$
are, respectively, 
the twist-3 distribution function for the transversely polarized nucleon and
the twist-3 fragmentation function for the pion.  
$\delta q(x)$ is the quark transversity
distribution of the nucleon, and $D(z)$ is the usual 
quark/gluon fragmentation function
for the pion.
In \cite{EKT06}, we focussed on 
the ``derivative term'' of the
soft-gluon-pole (SGP) contributions of (A) and (B) terms,
which is associated with the derivative of the corresponding 
twist-3 distribution function $G_F$ or fragmentation function $\widehat{E}_F$
and
is expected to be dominant in the large-$x_{bj}$ or large-$z_f$ region. 
Within this approximation,  
we also presented a simple estimate of the (A) and (B) contributions:
We fix the relevant nonperturbative functions involved in the (A) and (B) contributions
so that the similar twist-3 mechanism associated with those functions reproduces 
the data on 
$A_N$ in $pp^\uparrow\to\pi X$.
It turned out that the (B) term is negligible compared with 
the (A) term for SSA in SIDIS, even if we require
that each of the corresponding (A)- and (B)-type 
contributions can independently reproduce $A_N$ in $pp^\uparrow\to \pi X$.

A formalism of the twist-3 mechanism for SSA in the framework 
of collinear factorization
was first presented by Qiu and Sterman
for the direct photon production\,\cite{QS91}, and it was applied to various 
other processes\,\cite{QS99,KK00,Koike03,KQVY06,KK01,hyperon,EKT06,JQVY06,JQVY06DIS}. 
In this formalism, a coherent 
gluon field $A^\mu$ generated from quark-gluon correlation inside transversely 
polarized nucleon participates into the hard scattering subprocess. 
A systematic procedure for extracting the associated twist-3 effect
is the collinear expansion
of the relevant Feynman diagrams
in terms of the parton's transverse momenta $k_\perp$. 
Thus we have to perform the collinear expansion of a set of 
Feynman diagrams for the partonic hard scattering
with the extra gluon field $A^\mu$ attached.
On the other hand, gauge-invariant twist-3 quark-gluon
correlation functions appearing in the factorization formula of the cross section
are eventually expressed in terms of either the transverse components of the covariant derivative,
$D^\alpha=\partial^\alpha -ig A^\alpha$ with $\alpha=\perp$,
or those from the gluon
field strength tensor, $F^{\alpha +}= \partial^\alpha A^+ -\partial^+A^\alpha-ig[A^\alpha,A^+]$,
see (\ref{twist3distr}), (\ref{twist3D}) below.
Therefore, a crucial step for the calculation 
is to reorganize
the terms from the collinear expansion of 
the diagrams
in terms of the gauge-invariant combination corresponding to 
$D^\alpha$ or $F^{\alpha +}$
with $\alpha=\perp$
in the twist-3 accuracy.
In \cite{QS91}, the calculation was performed in the Feynman gauge, and
the combination
$\partial^\perp A^+$ 
resulting from the collinear expansion 
is identified as a part of
$F^{\perp +}=\partial^\perp A^+  - \partial^+A^\perp + \ldots$,
without any assurance about the gauge invariance of the procedure nor
the uniqueness of the obtained results. 
To justify this procedure,
it is necessary to show that ``$-\partial^+ A^\perp$'' also appears with 
exactly the same coefficient 
as that for $\partial^\perp A^+$.
Note that, in the Feynman gauge, the matrix element of the type 
$\la p\ S| \bar{\psi}\partial^\perp A^+\psi |p\ S \ra$ 
is equally important as $\la p\ S|  \bar{\psi}\partial^+A^\perp \psi |p\ S \ra$, 
and both matrix elements have to be treated as the same order in the 
collinear expansion. This is because, even though
$\la p\ S| \bar{\psi}A^+\psi |p\ S \ra \gg \la p\ S| \bar{\psi} A^\perp \psi |p\ S \ra$ 
in the Feynman gauge 
in a frame with $p^+ \gg p^- , p^\perp$, 
acting $\partial^\perp$ to the coherent gluon field in the former matrix 
element brings relative suppression 
compared with 
the latter with 
$\partial^+$ acting on the gluon field.
Accordingly, it is necessary to identify and keep the $\partial^+ A^\perp$ contribution
to perform the calculation in a consistent way, 
taking into account all terms in the collinear expansion up to the retained order,
which was missed in the literature.  
This eventually allows us to establish for the first time that
the corresponding factorization formula indeed holds for the twist-3 mechanism of SSA,
guaranteeing its gauge invariance and uniqueness.
Through this development,
we can also address 
the relations among possible different expressions for the cross section.

The purpose of writing this paper is twofold: First of all,  
we present all necessary ingredients to establish a systematic collinear-expansion
formalism for the twist-3 calculation
for SIDIS, $ep^\uparrow\to e\pi X$.
We will 
clarify the relation between different definitions of
twist-3 distributions, and identify and classify all pole contributions 
that can generate the interfering phase for the partonic subprocess.
Then we will pay particular attention to the consistency
relations for the hard part of 
the cross section
which are required from gauge invariance. Those relations express sufficient condition
to allow us to reorganize the next-to-leading terms in the collinear expansion in terms of 
the combination
$\partial^\perp A^+  - \partial^+A^\perp$ in a unique way, so that eventually  
justify the identification of $\partial^\perp A^+\to F^{\perp +}$ taken 
in \cite{QS91}.  
This is a nontrivial issue since the ``$\perp$'' components of a four-vector are treated 
in a completely different way 
from its ``$+$'' 
component in the collinear expansion.
We will next show that all kinds of pole contributions
for $ep^\uparrow\to e\pi X$ indeed satisfy the above-mentioned consistency relations, respectively.
Based on this development,
we derive the complete cross section formula
corresponding to the (A) term in (\ref{twist3}) including all pole contributions that produce SSA.
The obtained cross section formula 
is valid in all kinematic region for large $p_T$ pion production, 
and has a phenomenological relevance
for, e.g., the eRHIC experiment\,\cite{erhic}. 

The remainder of this paper is organized as follows:
In section 2, after recalling the basic properties 
of a complete set 
of the twist-3 quark-gluon correlation functions for the
transversely polarized nucleon and the kinematics of SIDIS,
we present our formalism of the twist-3 calculation of SSA in SIDIS.  
We keep all the subleading contributions of twist-3 from all diagrams for the cross section,
and show that they can be expressed in terms of the twist-3
gauge-invariant correlation functions introduced earlier
if the corresponding hard part satisfies certain constraints. 
The twist-2 cross section formula for the unpolarized SIDIS
is also given for completeness.  
In section 3, we derive the complete result for the factorization formula of the 
twist-3 cross section
corresponding to the (A) term in (\ref{twist3}).
To this end,
we shall prove that all kinds of the pole contributions
indeed satisfy the constraints discussed in section 2 by using the Ward identities.
We also provide an economic way of actual calculation, and 
discuss the relation between different expressions for the cross section.   
In section 4, we will present a numerical estimate
of SSA in SIDIS, using our cross section formula combined with a simple model for SGP functions.
We will include all terms from the SGP contributions,
the so-called ``derivative'' and ``non-derivative'' terms of the SGP functions, 
to see the importance of the latter contribution
which was not included in our previous study\,\cite{EKT06}.
Brief summary and conclusion are
presented in section 5.

\section{Formalism for the Twsit-3 mechanism in QCD
}
\subsection{A complete set of Twist-3 distribution functions}

We first recall from \cite{EKT06} the definition and the basic properties
of the twist-3 
distribution functions for the transversely polarized nucleon. 
They are characterized by the explicit participation of the
gluon field in the lightcone correlation functions for the nucleon. 
For the transversely polarized nucleon, it is known that there are two independent
twist-3 quark-gluon correlation functions\,\cite{KT99}. 
Actually, two types of different definitions for those functions
have been used in the literature, depending on 
whether the explicit gluon is introduced 
through the field strength tensor $F^{\alpha\beta}$ of the gluon
or the transverse components of the covariant derivative,
$D_\perp^\alpha=\partial_{\perp}^\alpha
-ig A_{\perp}^\alpha$. 
For the transversely polarized nucleon
with momentum $p^{\mu}$ and spin $S_\perp^\mu$,
they are given\,\cite{QS91,KT99} as the ``$F$-type'' functions,
\begin{eqnarray}
& &\int {d\lambda\over 2\pi}\int{d\mu\over 2\pi}
e^{i\lambda x_1}e^{i\mu(x_2-x_1)}
\langle p\ S_\perp |\bar{\psi}_j(0)[0,\mu n]
{gF^{\alpha\beta}(\mu n)n_\beta}[\mu n, \lambda n]
\psi_i(\lambda n)|p\ S_\perp \rangle\nonumber\\
& &\qquad=
{M_N\over 4} \left(\pslash\right)_{ij} 
\epsilon^{\alpha pnS_\perp}
{G_F(x_1,x_2)}+ 
i{M_N\over 4} \left(\gamma_5\pslash\right)_{ij} 
S_\perp^\alpha
{\GFt(x_1,x_2)}+ \cdots,
\label{twist3distr}
\end{eqnarray}
and the ``$D$-type'' functions,
\begin{eqnarray}
& &\int {d\lambda\over 2\pi}\int{d\mu\over 2\pi}
e^{i\lambda x_1}e^{i\mu(x_2-x_1)}
\langle p\ S_\perp |\bar{\psi}_j(0)[0,\mu n]
{D_\perp^{\alpha}(\mu n)}[\mu n, \lambda n]
\psi_i(\lambda n)|p\ S_\perp \rangle\nonumber\\
& &\qquad=
{M_N\over 4} \left(\pslash\right)_{ij} 
\epsilon^{\alpha pnS_\perp}
{G_D(x_1,x_2)}+ 
i{M_N\over 4} \left(\gamma_5\pslash\right)_{ij} 
S_\perp^\alpha
{\GDt(x_1,x_2)}+ \cdots,
\label{twist3D}
\end{eqnarray}
where $\psi$ is the quark field,
the nucleon's momentum $p$ can be regarded as
light-like ($p^2=0$) in the twist-3 accuracy, 
$n^\mu$ is the light-like vector 
($n^2=0$) with $p\cdot n=1$,
$\epsilon^{\alpha pnS_\perp}=\epsilon^{\alpha\lambda\mu\nu}p_\lambda
n_\mu S_{\perp \nu}$ with $\epsilon_{0123}=1$, 
and the spin vector satisfies $S_\perp^2 = -1$, $S_\perp \cdot p=S_\perp \cdot n=0$.  
 $[\mu n,\lambda n]={\rm P} \exp\left[ ig \int_{\lambda}^{\mu}dt n\cdot A(tn) \right]$ 
represents the gauge-link which guarantees gauge invariance of the
nonlocal lightcone operator,
and ``$\cdots$'' denotes twist-4 or higher-twist distributions. 
These four correlation functions, $G_F(x_1,x_2)$, $\GFt(x_1,x_2)$, 
$G_D(x_1,x_2)$, $\GDt(x_1,x_2)$, 
are defined as dimensionless by introducing the nucleon mass $M_N$ 
which is a natural scale for chiral-symmetry breaking.  
Each correlation function depends on the two variables $x_1$ and $x_2$, 
where $x_1$ and $x_2 -x_1$ are the fractions of the lightcone momentum 
carried by the quark and the gluon, respectively, which are outgoing from the nucleon.
From hermiticity, P- and T-invariance of QCD, these functions are real, and satisfy 
the symmetry properties
\beq
& &G_F(x_1,x_2)=G_F(x_2,x_1),\qquad\GFt(x_1,x_2)=-\GFt(x_2,x_1),
\label{symF}\\
& &G_D(x_1,x_2)=-G_D(x_2,x_1),\qquad\GDt(x_1,x_2)=\GDt(x_2,x_1).
\label{symD}
\eeq
The basis of twist-3 quark-gluon correlation functions for the
transversely polarized nucleon, defined as (\ref{twist3distr}) and (\ref{twist3D}),
is overcomplete.  
In \cite{EKT06}, the relation between the $F$-type functions of (\ref{twist3distr})
and the $D$-type functions of (\ref{twist3D})
was also
established using the QCD equations motion. They are connected by the relations
\beq
G_D(x_1,x_2)&=&P{1\over x_1-x_2}G_F(x_1,x_2),
\label{FDvector}\\
\GDt(x_1,x_2)&=&\delta(x_1-x_2)\gt(x_1)+
P{1\over x_1-x_2}\GFt(x_1,x_2), 
\label{FDrelation}
\eeq
where $P$ denotes the principal value, and $\gt(x_1 )$ can be expressed in terms of
the $F$-type distributions $G_F(x,y)$ and $\GFt(x,y)$, 
and the twist-2 quark helicity distribution $\Delta q(x)$ of the nucleon (see \cite{EKT06} for the
detail).
In the RHS of (\ref{FDvector}), the term proportional to
$\delta(x_1-x_2)$
vanishes due to the anti-symmetry of $G_D(x_1,x_2)$ as given in (\ref{symD}).   

It is convenient to choose the $F$-type functions
$\{G_F(x_1,x_2),\ \GFt (x_1,x_2)\}$ as a complete set of
twist-3 quark-gluon correlation functions for the
transversely polarized nucleon:
From (\ref{FDvector}) and (\ref{FDrelation}), we note 
that the $D$-type functions are more singular than
the $F$-type functions at the soft gluon point $x_1=x_2$, while
they are proportional to each other for $x_1\neq x_2$.  
As will be discussed in our analysis for $ep^\uparrow\to e\pi X$ below, 
the
cross section receives contributions from 
the ``soft gluon point'',  
$x_1=x_2$, of the twist-3 distribution; 
we assume
that there is no singularity in 
the $F$-type functions, in particular, 
that $G_F(x_1,x_2)$ is finite at $x_1=x_2$ ($\GFt(x_1,x_1)=0$, due to (\ref{symF})), 
and use
the 
$F$-type functions
to express those contributions in the cross section.
Other terms which receive contributions with $x_1\neq x_2$ are 
expressed in terms of either $D$-type or $F$-type functions without any subtlety. 

\subsection{Kinematics for $ep^\uparrow\to e\pi X$}

Here we summarize the kinematics for the SIDIS,
$e(\ell)+p(p, S_{\perp})\to e(\ell')+\pi(P_h)+X$.
\footnote{In this paper, we use different notation for the kinematic variables
compared with our previous paper\,\cite{EKT06}.}
(See refs.~\cite{MOS92,KN03} for the detail.)  
We have five independent Lorentz invariants, 
$S_{ep}, x_{bj},Q^2,z_f$, and $q_T^2$.  
The center of mass energy squared, $S_{ep}$, for the initial
electron and the proton is
\beq
S_{ep}=(p + \ell)^2 \simeq 2 p\cdot \ell \; ,
\eeq
ignoring masses. 
In the twist-3 accuracy, 
one can set the masses to be zero for all the particles in the initial and the final states.  
The conventional DIS variables are defined 
in terms of the virtual photon momentum $q=\ell-\ell'$ as
\beq
x_{bj}={Q^2\over 2p\cdot q} \; , \qquad Q^2 =-q^2=-(\ell-\ell')^2. 
\label{xbj}
\eeq
For the final-state pion, we introduce the scaling variable
\beq
z_f={p\cdot P_h\over p\cdot q} \; .
\eeq
Finally, we define the ``transverse'' component of $q$, which is orthogonal to
both $p$ and $P_h$:
\beq
q_t^\mu=q^\mu- {P_h\cdot q\over p\cdot P_h}p^\mu -
{p\cdot q\over p\cdot P_h}P_h^\mu \; .
\eeq
$q_t$ is a space-like vector, and we denote its magnitude by
\beq
q_T = \sqrt{-q_t^2} \; .
\eeq
To derive the cross section, we work in a frame where 
the
3-momenta $\vec{q}$ and $\vec{p}$ of the virtual photon and the initial proton are collinear, 
and we denote
the azimuthal angle
between the lepton plane and the hadron plane as $\phi$.  

For the actual calculation, 
it is convenient to 
specify the frame further.  
We work in the so-called {\it hadron frame}~\cite{MOS92},
which is the Breit frame of the virtual photon and the initial proton:
\beq
q^\mu &=& (0,0,0,-Q) \; ,\\
p^\mu &=& \left( {Q\over 2x_{bj}},0,0,{Q\over 2x_{bj}}\right) \; .
\eeq
In this frame the outgoing pion is taken to be in the $xz$ plane:
\beq
P_h^\mu = {z_f Q \over 2}\left( 1 + {q_T^2\over Q^2},{2 q_T\over Q},0,
{q_T^2\over Q^2}-1\right) \; .
\label{eq2.p_B}
\eeq
As one can see, the transverse momentum of the pion in this 
frame is given by $P_{hT}=z_f q_T$. This is true for any frame in which
the
3-momenta $\vec{q}$ and $\vec{p}$ 
are collinear.
The lepton momentum in this frame can be parameterized as
\beq
\ell^\mu={Q\over 2}\left( \cosh\psi,\sinh\psi\cos\phi,
\sinh\psi\sin\phi,-1\right) \; ,
\label{eq2.lepton}
\eeq
where
\beq
\cosh\psi = {2x_{bj}S_{ep}\over Q^2} -1 \; .
\label{eq2.cosh}
\eeq
We parameterize the transverse spin vector of the initial proton $S_\perp^\mu$
as
\beq
S^\mu_\perp = (0,\cos\Phi_S,\sin\Phi_S,0),
\label{phis}
\eeq
where $\Phi_S$ represents the azimuthal angle of $\vec{S}_\perp$ measured from
the hadron plane.
With the above definition, 
the cross section for $ep^\uparrow\to e\pi X$ can be expressed in terms of
$S_{ep}$, $x_{bj}$, $Q^2$, $z_f$, $q_T^2$, $\phi$ and $\Phi_S$ in the hadron
frame.  Note that $\phi$ and $\Phi_S$ are invariant under boosts in the 
$\vec{q}$-direction, so that the cross section presented below is
the same in any frame where $\vec{q}$ and $\vec{p}$ are collinear.  

\subsection{Collinear expansion and gauge invariance}
The differential cross section for
$ep^\uparrow\to e\pi X$ can be written as 
\beq
d\sigma={1\over 2S_{ep}}{d^3 \vec{P}_h\over (2\pi)^3 2P_h^0}{d^3\vec{\ell}'\over(2\pi)^3 
2\ell'^0}
{e^4\over q^4}L^{\mu\nu}(\ell,\ell')W_{\mu\nu}(p,q,P_h),
\label{dsigma}
\eeq
where 
\beq
L_{\mu\nu}(\ell,\ell')=2(\ell_\mu \ell'_\nu+\ell_\nu \ell'_\mu )-g_{\mu\nu}Q^2
\label{lepton}
\eeq
is the leptonic tensor for the unpolarized electron and
$W_{\mu\nu}$ is the hadronic tensor.
In the present study we are interested in the contribution from the twist-3 distribution for 
the initial proton
combined with the twist-2 unpolarized fragmentation function for the final pion
((A) term in (\ref{twist3})).  
Accordingly, it is straightforward to
factorize the fragmentation function from the hadronic tensor as
\beq
W_{\mu\nu}(p,q,P_h)=\sum_{j=q,g}\int{dz\over z^2}D_j(z) w^j_{\mu\nu}(p,q,{P_h\over z})\ ,
\label{Wmunu}
\eeq
where $D_j(z)$ ($j=q,\ g$) is the quark and gluon fragmentation functions for the pion, 
with $z$ being the momentum fraction.  
Since the calculational procedure is the same for the two terms in (\ref{Wmunu}), 
we consider the case for the quark fragmentation in detail and
omit the index $j$ from $w_{\mu\nu}^j$ below.  
To extract the twist-3 effect in $w_{\mu\nu}$
contributing to single-spin-dependent cross section, 
one needs to
analyze the diagrams shown in Fig. 1 for the hadronic tensor.
To this end, 
we work in the Feynman gauge.  
In Fig.~1, the partons from the nucleon, which is represented  
by the lower blob, undergoes the hard interaction with the virtual photon in the middle blob,
followed by the fragmentation into the pion as represented by the upper blob.   
Note that the fragmentation function corresponding to the upper blob has already been 
factorized as (\ref{Wmunu}).
Corresponding to (a) and (b) of Fig. 1, where the momenta for the parton lines connecting the blobs
are assigned, respectively, 
$w_{\mu\nu}$ can be written as the sum of the two terms:
\beq
w_{\mu\nu}(p,q,{P_h\over z})&=&
\int{d^4k\over (2\pi)^4}{\rm Tr}\left[ 
M^{(0)}(k)S^{(0)}(k,q,{P_h\over z})\right]
\nonumber\\
&+&\int{d^4k_1\over (2\pi)^4} \int{d^4k_2\over (2\pi)^4} 
{\rm Tr} 
\left[  M^{(1)\sigma} (k_1,k_2)S^{(1)}_\sigma (k_1,k_2,q,{P_h\over z})\right], 
\label{w-tensor}
\eeq
where the superscripts ``$(0)$'' and ``$(1)$'' indicate the number of coherent gluon line
connecting the lower and middle blobs.
$M^{(0)}(k)$ and $M^{(1)\sigma}(k_1,k_2)$ represent the lower part of (a) and (b) of Fig. 1, 
respectively, and they are given as the nucleon matrix elements: 
\beq
M^{(0)}_{ij}(k)&=&\int\,d^4\xi\, e^{ik\xi}\la p\ S_\perp|\bar{\psi}_j(0)\psi_i(\xi)|p\ S_\perp\ra,
\label{m00}
\\
M^{(1)\sigma}_{ij}(k_1,k_2)&=&\int\,d^4\xi\int\,d^4\eta\, e^{ik_1\xi}e^{i(k_2-k_1)\eta}
\la p\ S_\perp|\bar{\psi}_j(0)gA^\sigma(\eta)\psi_i(\xi)|p\ S_\perp\ra,
\label{m1sigma}
\eeq
where the spinor indices $i$ and $j$ are shown explicitly;
$S^{(0)}(k,q,{P_h\over z})$ and $S^{(1)}_\sigma (k_1,k_2,q,{P_h\over z})$ 
represent the corresponding hard parts
with the Lorentz indices $\mu$ and $\nu$ suppressed here and below
for simplicity,
and 
are matrices in spinor space. 
${\rm Tr}[\cdots ]$ indicates the trace over Dirac-spinor indices
while the color trace is implicit. 
Since the twist-2 unpolarized fragmentation
functions $D_j (z)$ in (\ref{Wmunu}) are chiral-even quantities,
only chiral-even Dirac matrix structures in 
$M^{(0)}(k)$ and $M^{(1)\sigma}(k_1,k_2)$ give nonvanishing contribution to ${\rm Tr}[\cdots ]$.

\begin{figure}[t!]
\begin{center}
\epsfig{figure=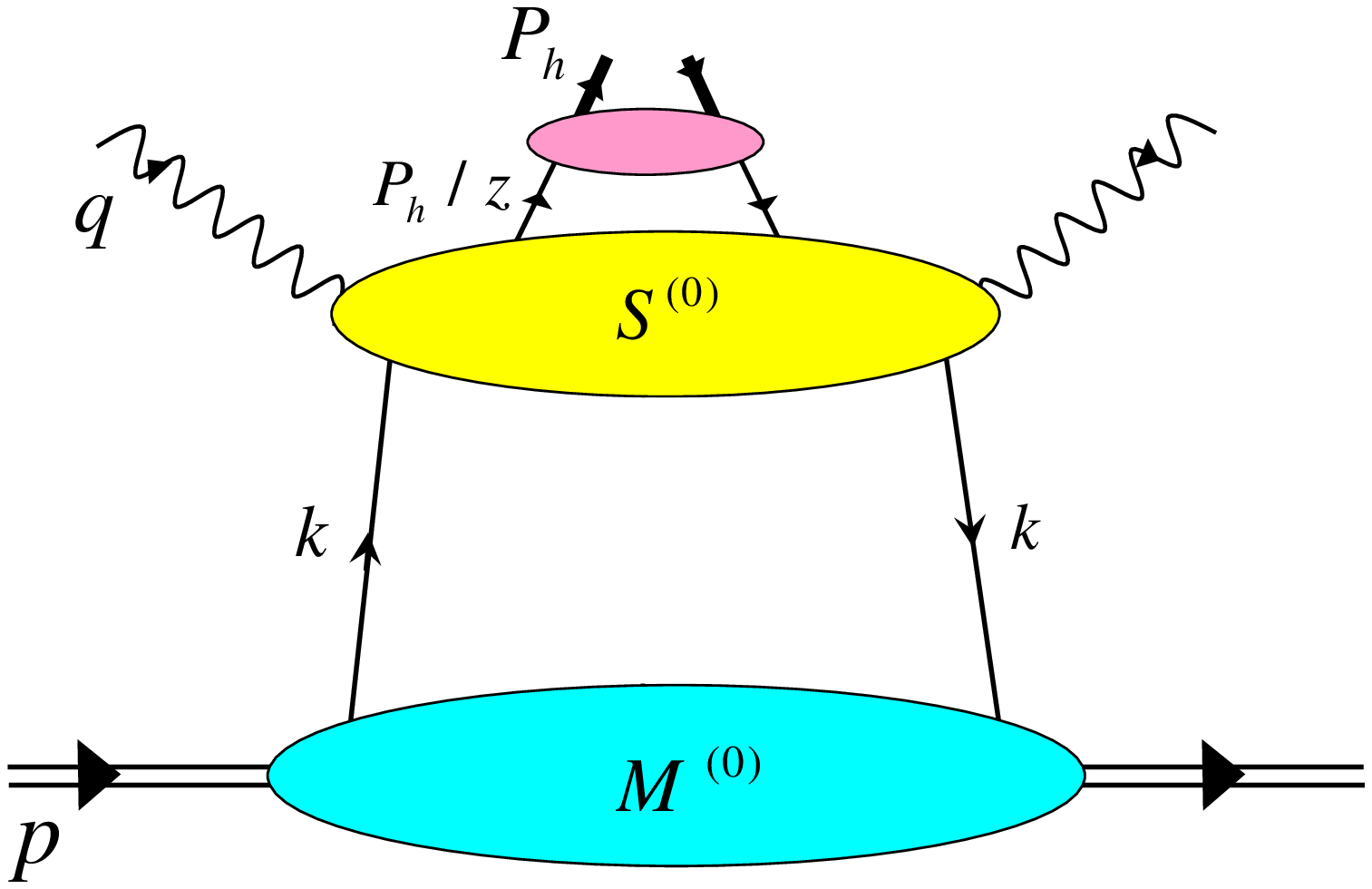,width=0.4\textwidth,clip=}
\hspace{22mm}
\epsfig{figure=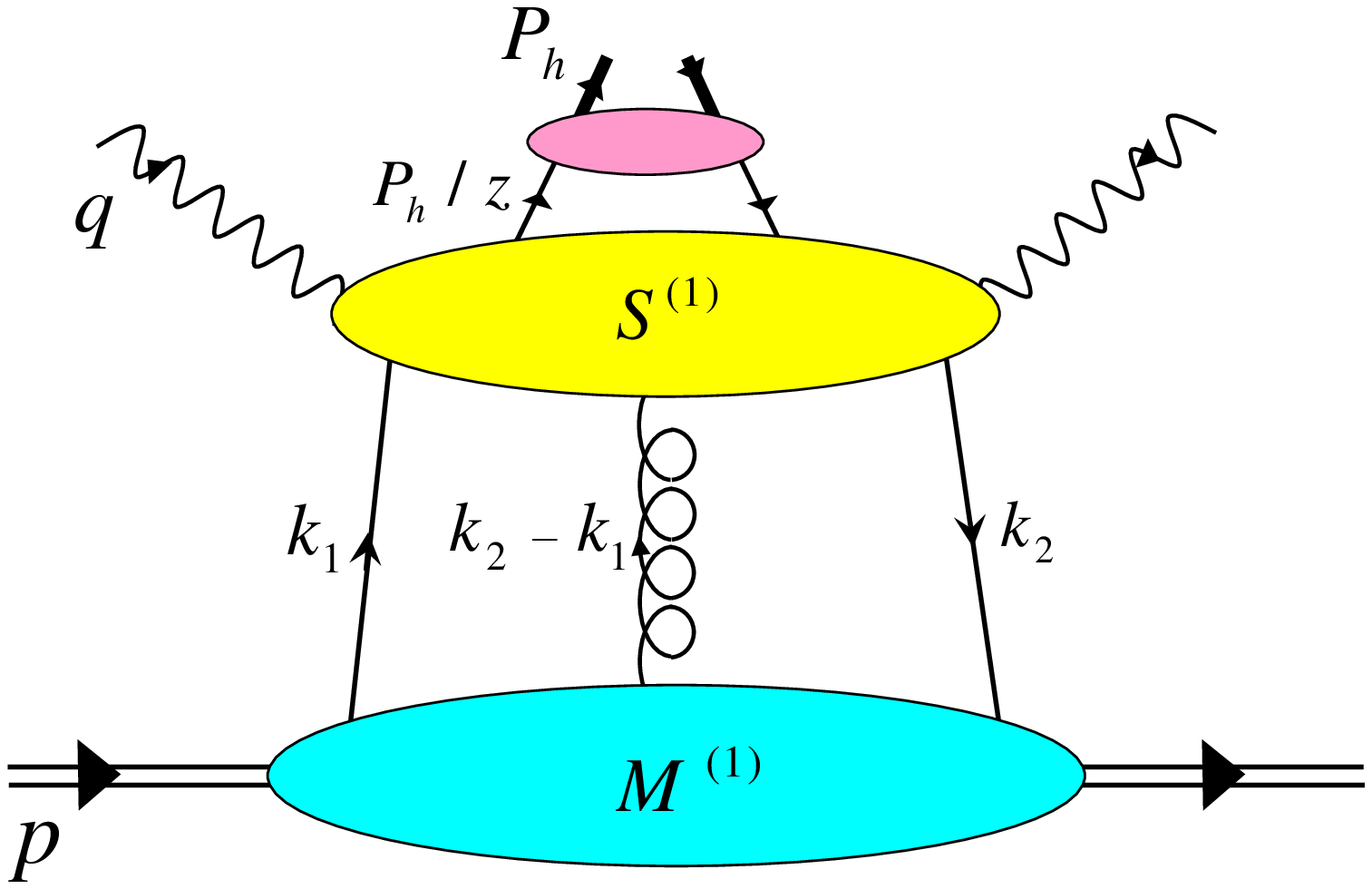,width=0.4\textwidth,clip=}
\end{center}
\vspace{-0.5cm}
\hspace{3.4cm}(a)
\hspace{8.4cm}(b)
\caption{Generic diagrams for the hadronic tensor of $ep^\uparrow\to e\pi X$,
decomposed into the three blobs as nucleon matrix element (lower), pion matrix element (upper),
and partonic hard scattering by the virtual photon (middle). The first two terms, (a) and (b),  
in the expansion by the number of partons connecting the middle and lower blobs
are relevant to the twist-3 effect induced by the nucleon.
\label{fig1}}
\vspace*{0.cm}
\end{figure}

In the leading order in QCD perturbation theory, the hard blob for $S^{(0)}(k,q,{P_h\over z})$ 
in Fig. 1 (a)
stands for a set of cut Feynman diagrams for the partonic Born subprocess with a hard gluon 
or quark in the final state,
while the additional, coherent gluon participates 
in those partonic processes for $S^{(1)}_\sigma (k_1,k_2,q,{P_h\over z})$ in Fig. 1 (b). 
A standard and systematic procedure 
to identify 
the twist-3 contributions
to $w_{\mu\nu}$ of (\ref{w-tensor}) utilizes the collinear expansion 
of the hard parts 
$S^{(0)}(k,q,{P_h\over z})$ and $S^{(1)}_\sigma (k_1,k_2,q,{P_h\over z})$ 
with respect to the partonic momenta
$k$ and $k_{1,2}$ around the momentum $p$ of the parent nucleon.  
This reads for $S^{(0)}(k,q,{P_h\over z})$
\beq
S^{(0)}(k,q,P_h/z)=S^{(0)}(xp,q,P_h/z)+
\left.{\partial S^{(0)}(k,q,P_h/z)\over \partial k^\alpha}\right|_{k=xp}
\omega^\alpha_{\ \beta}k^\beta+\cdots,
\label{colex}
\eeq
where $x=k\cdot n$, 
$\omega^{\alpha\beta}\equiv g^{\alpha\beta}-p^\alpha n^\beta$, and ``$\cdots$'' denote
the terms of $O\left((k_\perp)^2 \right)$ which are  
twist-4 or higher.  
Substituting
this expansion,
the first term in the RHS of (\ref{w-tensor}) becomes
\beq
\int\;dx\;{\rm Tr} \left[ M^{(0)}(x)S^{(0)}(xp,q,P_h/z)\right]+
\int\;dx\;{\rm Tr}\left[ i\omega^{\alpha}_{\ \beta}M^{(0)\beta}_\partial (x)
\left.{\partial S^{(0)}(k,q,P_h/z)\over \partial k^\alpha}\right|_{k=xp}\right],
\label{collinear0}
\eeq
where the integration over $k^-$ and $k^\perp$ has been performed, with 
$\omega^\alpha_{\ \beta}k^\beta$ of (\ref{colex}) transformed into the derivative 
$i\omega^\alpha_{\ \beta}\partial^\beta$ 
on the quark field of (\ref{m00})
by partial integration,  
and 
the relevant nucleon matrix elements have now become the lightcone correlation functions: 
\beq
M^{(0)}_{ij}(x)&=&\int{d\lambda\over 2\pi}e^{i\lambda x}
\la p\ S_\perp|\bar{\psi}_j(0)\psi_i(\lambda n)|p\ S_\perp\ra,
\label{eq24}
\\
M^{(0)\alpha}_{\partial\ ij}(x)&=&\int{d\lambda\over 2\pi}e^{i\lambda x}
\la p\ S_\perp|\bar{\psi}_j(0)\partial^\alpha\psi_i(\lambda n)|p\ S_\perp\ra.
\label{eq25}
\eeq
Here the matrix elements $M^{(0)}(x)$ and $M^{(0)\alpha}_{\partial}(x)$ 
for the transversely polarized nucleon
can be decomposed
into several independent Dirac matrix structures 
based on Lorentz, P- and T-invariance. 
Among them, 
we only need to consider the 
chiral-even Dirac matrix structures as noted above.
We find $M^{(0)}(x) = \gamma_5\Sslash_\perp g_T(x)/2 + \ldots$ with 
the chiral-even twist-3 quark distribution $g_T(x)$\,\cite{JJ92}, 
up to irrelevant (chiral-odd or twist-4) terms,
but the first term in (\ref{collinear0}) with the substitution $M^{(0)}(x) 
\rightarrow \gamma_5\Sslash_\perp g_T(x)/2$  
does not contribute to the cross section:
The spinor trace ${\rm Tr}[\cdots ]$ involving $\gamma_5$ 
produces the factor $i$ to the first term in (\ref{collinear0}) and the result
is contracted 
with 
the real tensor $L_{\mu\nu}$ of (\ref{lepton}) to derive the contribution to (\ref{dsigma});
but this yields a real quantity for the cross section only when another factor $i$ is provided by 
the hard part $S^{(0)}(xp,q,P_h/z)$, which is impossible for the Born subprocess.   
Similarly, it is straightforward to see that the second term of (\ref{collinear0}) does not contribute
to the cross section to twist-3 accuracy, noting that the second term of (\ref{collinear0}) 
contains the factor $i$ and 
considering a similar decomposition of $\omega^{\alpha}_{\ \beta}M^{(0)\beta}_{\partial}(x)$ 
into Dirac matrix structures as in (\ref{twist3D}).

Therefore the twist-3 cross section for $ep^\uparrow\to e\pi X$ arises solely from
the second term of (\ref{w-tensor}).  To analyze it, we first note
that the matrix element $M^{(1)\sigma=\perp}$ of (\ref{m1sigma}) is suppressed by a power of $1/p^+$ 
compared to
$M^{(1)\sigma=+}$, and gives subleading contribution to $w_{\mu\nu}$. 
Accordingly, in the twist-3 accuracy, 
collinear expansion is necessary only for
$S^{(1)}_+ =S^{(1)}_\sigma p^\sigma /p^+$ which couples with $M^{(1)+}$ in (\ref{w-tensor}).
We define
\beq
S^{(1)}(k_1,k_2,q,P_h/z)\equiv
S^{(1)}_\sigma(k_1,k_2,q,P_h/z) p^\sigma.
\eeq
Collinear expansion for $S^{(1)}(k_1,k_2,q,P_h/z)$ with respect to $k_{1,2}$ around
$p$ gives 
\beq
S^{(1)}(k_1,k_2,q,P_h/z)
&=&S^{(1)}(x_1p,x_2p,q,P_h/z)+
\left.{\partial S^{(1)}(k_1,k_2,q,P_h/z)\over \partial k_1^\alpha}\right|_{k_i=x_ip}
\omega^\alpha_{\ \beta}k_1^\beta\nonumber\\
&+&
\left.{\partial S^{(1)}(k_1,k_2,q,P_h/z)\over \partial k_2^\alpha}\right|_{k_i=x_ip}
\omega^\alpha_{\ \beta}k_2^\beta+\cdots,
\label{collinear1}
\eeq
where $x_i = k_i \cdot n$ with $i=1, 2$.
Using (\ref{collinear1}) in the second term of (\ref{w-tensor}), one obtains 
\beq
w_{\mu\nu} = & &\int\;dx_1\int\;dx_2{\rm Tr}\left[ M^{(1)}_\sigma (x_1,x_2)n^\sigma 
S^{(1)}(x_1p,x_2p,q,P_h/z)\right]\nonumber\\
& &+
\int\;dx_1\int\;dx_2{\rm Tr}\left[ \omega^\alpha_{\ \beta}M^{(1)\beta} (x_1,x_2)
S_\alpha^{(1)}(x_1p,x_2p,q,P_h/z)\right]\nonumber\\
& &+
\int\;dx_1\int\;dx_2{\rm Tr}\left[ i\omega^\alpha_{\ \beta}M^{(1)\beta}_{\partial 1} (x_1,x_2)
\left.{\partial S^{(1)}(k_1,k_2,q,P_h/z)
\over \partial k_1^\alpha}\right|_{k_i=x_ip}
\right]\nonumber\\
& &+
\int\;dx_1\int\;dx_2{\rm Tr}\left[ i\omega^\alpha_{\ \beta}M^{(1)\beta}_{\partial 2} (x_1,x_2)
\left.{\partial S^{(1)}(k_1,k_2,q,P_h/z)
\over \partial k_2^\alpha}\right|_{k_i=x_ip}
\right],
\label{w-tensor1}
\eeq
up to twist-3 contributions, and, similarly to (\ref{eq24}) and (\ref{eq25}) above,
the relevant nucleon matrix elements have now become the correlation functions on the
lightcone:
\beq
M^{(1)\sigma}(x_1,x_2)&=&
\int{d\lambda\over 2\pi}
\int{d\mu\over 2\pi}\;e^{i\lambda x_1}e^{i\mu(x_2-x_1)}
\la p\ S_\perp|\bar{\psi}(0)gA^\sigma(\mu n) \psi(\lambda n) |p\ S_\perp\ra,\\
M^{(1)\beta}_{\partial 1}(x_1,x_2)&=&
\int{d\lambda\over 2\pi}
\int{d\mu\over 2\pi}\;e^{i\lambda x_1}e^{i\mu(x_2-x_1)}
\la p\ S_\perp|\bar{\psi}(0)gA^\sigma(\mu n)n_\sigma\partial^\beta \psi(\lambda n) |p\ S_\perp\ra,
\\
M^{(1)\beta}_{\partial 2}(x_1,x_2)&=&
\int{d\lambda\over 2\pi}
\int{d\mu\over 2\pi}\;e^{i\lambda x_1}e^{i\mu(x_2-x_1)}
\la p\ S_\perp|\bar{\psi}(0)\left\{
g\left(\partial^\beta A^\sigma(\mu n)n_\sigma\right) \psi(\lambda n)\right.\nonumber\\
& &\qquad\qquad\left.+ gA^\sigma(\mu n)n_\sigma\partial^\beta \psi(\lambda n)\right\} |p\ S_\perp\ra.
\eeq
The first term of (\ref{w-tensor1})
does not contribute to the single-spin-dependent cross section
by the same reason as 
the first term of (\ref{collinear0}): Actually the former is
absorbed into the latter, providing the $O(g)$ contribution of the gauge-link operator 
to be inserted in-between the quark fields in $M^{(0)}(x)$
of (\ref{eq24}), as can be seen straightforwardly using the Ward identity for the coupling
of the scalar polarized, coherent gluon field $n\cdot A$ into the corresponding hard part
(see also \cite{qqss}),
\begin{equation}
S^{(1)}(x_1p,x_2p, q,P_h/z) =- \frac{S^{(0)}(x_2 p, q,P_h/z)}{x_1 - x_2 + i\epsilon} 
- \frac{S^{(0)}(x_1 p, q,P_h/z)}{x_2 - x_1 - i\varepsilon}\ .
\label{ward}
\end{equation}
We rewrite the remaining terms of (\ref{w-tensor1}) as
\beq
w_{\mu\nu}& &=\int\;dx_1\int\;dx_2{\rm Tr}\left[ \omega^\alpha_{\ \beta}M^{(1)\beta} (x_1,x_2)
S_\alpha^{(1)}(x_1p,x_2p,q,P_h/z)\right]\nonumber\\
& &+
\int\;dx_1\int\;dx_2{\rm Tr}\left[ i\omega^\alpha_{\ \beta}
\left(
M^{(1)\beta}_{\partial 2} (x_1,x_2)-
M^{(1)\beta}_{\partial 1} (x_1,x_2)
\right)
\left.{\partial S^{(1)}(k_1,k_2,q,P_h/z)\over \partial k_2^\alpha}\right|_{k_i =x_i p}
\right]\nonumber\\
& &+
\int\;dx_1\int\;dx_2{\rm Tr}\left[ i\omega^\alpha_{\ \beta}M^{(1)\beta}_{\partial 1} (x_1,x_2)
\right.\nonumber\\
& &\qquad\times\left.\left(
\left.{\partial S^{(1)}(k_1,k_2,q,P_h/z)
\over \partial k_1^\alpha}\right|_{k_i=x_ip}
+\left.{\partial S^{(1)}(k_1,k_2,q,P_h/z)
\over \partial k_2^\alpha}\right|_{k_i=x_ip}
\right)
\right], 
\label{w-twist3}
\eeq
and further rewrite $M^{(1)\beta}_{\partial 2} -
M^{(1)\beta}_{\partial 1}$ in the second term on the RHS as
\beq
M^{(1)\beta}_{\partial 2} (x_1,x_2)-
M^{(1)\beta}_{\partial 1} (x_1,x_2)
=M^{(1)\beta}_{F} (x_1,x_2)+
\widetilde{M}^{(1)\beta} (x_1,x_2),
\label{M12}
\eeq
with
\beq
M^{(1)\beta}_{F}(x_1,x_2)&=&
\int{d\lambda\over 2\pi}
\int{d\mu\over 2\pi}\;e^{i\lambda x_1}e^{i\mu(x_2-x_1)}
\la p\ S_\perp|\bar{\psi}(0)\nonumber\\
& &\qquad\qquad\times g\left\{\left(\partial^\beta A^\sigma(\mu n)\right)-
\left(\partial^\sigma A^\beta(\mu n)\right)\right\}n_\sigma
\psi(\lambda n) |p\ S_\perp\ra,
\label{MF}\\
\widetilde{M}^{(1)\beta}(x_1,x_2)&=&
\int{d\lambda\over 2\pi}
\int{d\mu\over 2\pi}\;e^{i\lambda x_1}e^{i\mu(x_2-x_1)}
\la p\ S_\perp|\bar{\psi}(0)gn_\sigma  \left(\partial^\sigma A^\beta(\mu n)\right) \psi(\lambda n) 
|p\ S_\perp\ra\nonumber\\
&=&-i(x_2-x_1)M^{(1)\beta}(x_1,x_2),
\label{Mtilde}
\eeq
where the second equality of (\ref{Mtilde}) is obtained by partial integration. 
One should note the two terms in (\ref{MF})
contribute in the same
order with respect to the power of the large scale $p^+$: 
Even though, schematically, in the power counting for $p^+$
\beq
\la \bar{\psi}A^+\psi\ra \gg \la \bar{\psi}A^\perp\psi\ra, 
\label{inequality1}
\eeq
by one power of $p^+$, 
one has 
\beq
\la \bar{\psi}(\partial^\perp A^+)\psi\ra \sim \la \bar{\psi}(\partial^+A^\perp)\psi\ra,
\label{inequality2}
\eeq
since $\partial^\perp$ acting on the coherent gluon field
causes 
relative power-suppression $\sim 1/p^+$
compared with $\partial^+$.
Using the relation (\ref{M12}) and (\ref{Mtilde})
in the second term of (\ref{w-twist3}),
one obtains 
\beq
w_{\mu\nu}& &
=\int\;dx_1\int\;dx_2{\rm Tr}\left[ i\omega^\alpha_{\ \beta}
M^{(1)\beta}_{F} (x_1,x_2)
\left. {\partial S^{(1)}(k_1,k_2,q,P_h/z)
\over \partial k_2^\alpha}\right|_{k_i=x_ip}
\right]\nonumber\\
& &
+\int\;dx_1\int\;dx_2{\rm Tr}\left[ 
\omega^\alpha_{\ \beta} M^{(1)\beta} (x_1,x_2)
\left\{
\left.(x_2-x_1){\partial S^{(1)}(k_1,k_2,q,P_h/z)
\over \partial k_2^\alpha}\right|_{k_i=x_ip}
\right.\right.\nonumber\\
& &\qquad\qquad\qquad\qquad\left.\left. +S_\alpha^{(1)}(x_1p,x_2p,q,P_h/z)
\right\}
\right]\nonumber\\
& &+
\int\;dx_1\int\;dx_2{\rm Tr}\left[ i\omega^\alpha_{\ \beta}M^{(1)\beta}_{\partial 1} (x_1,x_2)
\right.\nonumber\\
& &\qquad\times\left.\left(
\left.{\partial S^{(1)}(k_1,k_2,q,P_h/z)
\over \partial k_1^\alpha}\right|_{k_i=x_ip}
+\left.{\partial S^{(1)}(k_1,k_2,q,P_h/z)
\over \partial k_2^\alpha}\right|_{k_i=x_ip}
\right)
\right].
\label{w-twist3-f}
\eeq
In our analysis with a single coherent gluon connecting the hard part and the soft (nucleon)
part as in Fig. 1 (b), $M_F^{(1)\beta}(x_1 , x_2 )$ in the first term in the RHS of 
(\ref{w-twist3-f}) can be identified as
the gauge-invariant correlation function defined as (\ref{twist3distr})
with the gluon field strength tensor (see (\ref{MF})). 
On the other hand, the other terms in (\ref{w-twist3-f}) involve the matrix elements
that are not directly related to (\ref{twist3distr}).
Accordingly, 
if those terms
vanish,
the first term of (\ref{w-twist3-f}) gives 
the
cross section in terms of the $F$-type functions.
In order that this is the case, 
we may simply require that the hard parts for each of the second and third terms of 
(\ref{w-twist3-f}) vanish 
separately as
\beq
& &(x_2-x_1)\left.{\partial S^{(1)}(k_1,k_2,q,P_h/z)
\over \partial k_2^\alpha}\right|_{k_i=x_ip}
+S_\alpha^{(1)}(x_1p,x_2p,q,P_h/z)=0,
\label{consistency1}\\
& &\left.{\partial S^{(1)}(k_1,k_2,q,P_h/z)
\over \partial k_1^\alpha}\right|_{k_i=x_ip}
+\left.{\partial S^{(1)}(k_1,k_2,q,P_h/z)
\over \partial k_2^\alpha}\right|_{k_i=x_ip}=0,
\label{consistency2}
\eeq
for $\alpha = \perp$, up to the twist-4 corrections.
These constitute the sufficient condition to obtain the cross section formula
in terms of the gauge-invariant correlation functions. 
In Section~3, we will see that these relations are indeed satisfied
by the hard part for the partonic subprocess relevant to SSA.

A comment on the previous works
is in order regarding the relation (\ref{consistency1}).  
A formalism using the Feynman gauge for the twist-3 calculation of SSA
was 
first presented in \cite{QS91}. 
In \cite{QS91} 
and all the subsequent works\,\cite{QS99,KK00,Koike03,KQVY06,KK01,hyperon,EKT06,JQVY06,JQVY06DIS}, 
the identification of $\partial^\perp A^+\to F^{\perp+}$ in the matrix element
was done in the course of the collinear expansion
ignoring other terms in $F^{\perp +}$.
In \cite{LQS94}, the relation (\ref{inequality1}) was correctly recognized (see
the argument following eq. (86) of \cite{LQS94})
but the relation (\ref{inequality2}) was not noticed and  
the $\partial^+ A^\perp$ term in $F^{\perp +}$ was discarded. 
(see eqs. (27a) and (27b) of \cite{LQS94}).  
Clearly this procedure is not justified 
as was emphasized around (\ref{inequality1}) and (\ref{inequality2}).
It is this fact that forces us to reorganize the first and second terms of (\ref{w-twist3})
into those of (\ref{w-twist3-f}) and require eventually (\ref{consistency1}).
Therefore, we emphasize the condition (\ref{consistency1}) is crucial 
to justify the above identification, $\partial^\perp A^+\to F^{\perp+}$.

\subsection{Pole contribution to the cross section}
As has been known from the previous studies\,\cite{ET82,QS91}, 
the single-spin-dependent cross section occurs
as pole contributions of the internal propagators
in the partonic hard cross section.
This can be seen as follows: 
Since the leptonic tensor (\ref{lepton}) for the unpolarized lepton
is real,
the cross section receives the contributions
from the real part 
of the hadronic tensor $W_{\mu\nu}$ of (\ref{Wmunu}).
When (\ref{consistency1}) and (\ref{consistency2}) are satisfied,
the hadronic tensor is given by the first term of
(\ref{w-twist3-f}) which is proportional to the factor $i$.
Substituting (\ref{twist3distr}) into this term, 
we immediately see that the trace ${\rm Tr}[\cdots ]$ of the relevant Dirac matrices,
combined with $\pslash$ and $i\gamma_5\pslash$ of the two terms of (\ref{twist3distr}),
produces a real factor, 
so that the hadronic tensor receives the real contributions
only from 
the imaginary part
of $\partial S^{(1)}/\partial k_2^\alpha|_{k_i=x_ip}$. 
This requires to pick up 
imaginary part from a pole 
of an internal propagator in $\partial S^{(1)}/\partial k_2^\alpha|_{k_i=x_ip}$.

Analyzing the cut diagrams for the hard blob
in Fig. 1(b),
which represents the partonic Born subprocess with the additional, coherent gluon attached,
it is easy to see that, when the momenta of ``external'' partons are in the collinear 
configuration (i.e. $k_{1,2}=x_{1,2}p$),
some internal lines can go on shell, with the propagators reducing to 
one of the following three factors:
\beq
& &
({\rm I})\quad
{1\over x_1-x_{bj}+i\varepsilon}=
P{1\over x_1-x_{bj}}-i\pi\delta(x_1-x_{bj}),\nonumber\\
& &\qquad\qquad\qquad{\rm or}\quad{1\over x_2-x_{bj}-i\varepsilon}=
P{1\over x_2-x_{bj}}+i\pi\delta(x_1-x_{bj}),
\label{HPprop}\\
& &
({\rm II})\quad
{1\over x_1-i\varepsilon}=P{1\over x_1}+i\pi\delta(x_1),\quad{\rm or}
\quad{1\over x_2+i\varepsilon}=P{1\over x_2}-i\pi\delta(x_2),
\label{SFPprop}\\
& &
({\rm III})\quad
{1\over x_1-x_2+i\varepsilon}=P{1\over x_1-x_2}-i\pi\delta(x_1-x_2).
\label{SGPprop}
\eeq
These (I), (II), and (III) produce the pole contributions called
hard pole (HP)\,\cite{Guo98}, soft-fermion pole (SFP)\,\cite{QS91} and
soft-gluon pole (SGP)\,\cite{QS91}, respectively, when integrated over the parton momentum fraction.  
For $\partial S^{(1)}/\partial k_2^\alpha|_{k_i=x_ip}$ of (\ref{w-twist3-f}),
the derivative of the internal propagators may contribute as double pole as well as single pole,
but the position of the double pole is the same as that of (\ref{HPprop})-(\ref{SGPprop}).
In Section~3 we shall calculate all pole contributions to the cross section.

\subsection{Unpolarized cross section}

Here we present a brief description of the calculational procedure for SIDIS.
See \cite{MOS92,KN03} for the detail.  
Following \cite{MOS92}, 
we introduce the following 4 vectors which are orthogonal to
each other\,\cite{MOS92}:
\beq
T^\mu &=&{1\over Q}\left(q^\mu + 2x_{bj}p^\mu\right),\nonumber\\
X^\mu &=&{1\over q_T}\left\{ {P_h^\mu \over z_f} -q^\mu -\left(
1+{q_T^2\over Q^2}\right)x_{bj} p^\mu\right\},\nonumber\\
Y^\mu &=& \epsilon^{\mu\nu\rho\sigma}Z_\nu X_\rho T_\sigma ,\nonumber\\
Z^\mu &=& -{q^\mu \over Q}.
\eeq
These vectors become
$T^\mu=(1,0,0,0)$, $X^\mu=(0,1,0,0)$, $Y^\mu=(0,0,1,0)$, $Z^\mu=(0,0,0,1)$
in the hadron frame.  
Note that $T$, $X$ and $Z$ are vectors, while
$Y$ is an axial vector.
Since $W^{\mu\nu}$ satisfies the current conservation,
$q_\mu W^{\mu\nu} =q_\nu W^{\mu\nu}=0$, 
$W^{\mu\nu}$ can be expanded in terms of
9 independent tensors, for which one can employ the following:
\beq
{\cal V}_1^{\mu\nu}&=&X^\mu X^\nu + Y^\mu Y^\nu,\quad
{\cal V}_2^{\mu\nu}=g^{\mu\nu} + Z^\mu Z^\nu,\quad
{\cal V}_3^{\mu\nu}=T^\mu X^\nu + X^\mu T^\nu,\nonumber\\
{\cal V}_4^{\mu\nu}&=&X^\mu X^\nu - Y^\mu Y^\nu,\quad
{\cal V}_5^{\mu\nu}= i(T^\mu X^\nu - X^\mu T^\nu),\quad
{\cal V}_6^{\mu\nu}=i(X^\mu Y^\nu - Y^\mu X^\nu),\nonumber\\
{\cal V}_7^{\mu\nu}&=&i(T^\mu Y^\nu - Y^\mu T^\nu),\quad
{\cal V}_8^{\mu\nu}=T^\mu Y^\nu + Y^\mu T^\nu,\quad
{\cal V}_9^{\mu\nu}=X^\mu Y^\nu + Y^\mu X^\nu.
\eeq
For both unpolarized twist-2 cross section and single-spin-dependent twist-3 cross section,
the corresponding hadronic tensor $W^{\mu\nu}$ is symmetric. In addition, for both cases,
pseudo-tensor ${\cal V}_{8,9}$ do not contribute to the expansion of
$W^{\mu\nu}$, so that $W^{\mu\nu}$ can be expanded in terms of ${\cal V}_{i}$ ($i=1,\cdots,4$).  
The latter fact can be seen by explicit calculation, 
or by inspection: For the single-spin-dependent case,
substituting (\ref{twist3distr}) into the first term of 
(\ref{w-twist3-f}),
the trace ${\rm Tr}[\cdots ]$ of the relevant Dirac matrices
gives the factor $\epsilon^{P_hS_\perp pn} \propto \vec{S}_\perp \cdot (\vec{p}\times 
\vec{P}_{h})$ characteristic
of SSA,
which behaves as scalar under parity transformation.

The expansion coefficients of $W^{\mu\nu}$ for these tensors
are easily obtained by using the inverse tensors $\widetilde{{\cal V}}_k$
for ${\cal V}_k$ as
\beq
{W}^{\mu\nu} =\sum_{k=1}^4 {\cal V}_k^{\mu\nu} \left[W_{\rho\sigma}
\widetilde{{\cal V}}_k^{\rho\sigma}\right],
\label{eq2.expansion}
\eeq
where
\beq
& &\widetilde{{\cal V}}_1^{\mu\nu}={1\over 2}(2T^\mu T^\nu
+X^\mu X^\nu + Y^\mu Y^\nu),\qquad
\widetilde{{\cal V}}_2^{\mu\nu}=T^\mu T^\nu,\nonumber\\
& &\widetilde{{\cal V}}_3^{\mu\nu}=-{1\over 2}(T^\mu X^\nu + X^\mu T^\nu),\qquad
\widetilde{{\cal V}}_4^{\mu\nu}={1\over 2}(X^\mu X^\nu - Y^\mu Y^\nu).  
\eeq
With these ${\cal V}^{\mu\nu}_k$, we can decompose the $\phi$-dependence of
the cross sections by the contraction with $L^{\mu\nu}$ in (\ref{lepton}).
Define ${\cal A}_k$ ($k=1,\cdots,4$) as
\beq
{\cal A}_k={1\over Q^2}L_{\mu\nu}{\cal V}_k^{\mu\nu}\ ,
\eeq
then one obtains
\beq
{\cal A}_1&=&1+\cosh^2\psi,\nonumber\\
{\cal A}_2&=&-2,\nonumber\\
{\cal A}_3&=&-\cos\phi\sinh 2\psi,\nonumber\\
{\cal A}_4&=&\cos 2\phi\sinh^2\psi.
\label{Ak}
\eeq

With this method, the unpolarized cross section of twist-2 for $ep \to e\pi X$ was 
derived some time ago\,\cite{Mendez78,MOS92,KN03}. 
It reads
\beq
& &{d^5\sigma\over dx_{bj} dQ^2 dz_f dq_T^2 d\phi}
={\alpha_{em}^2 \alpha_S \over 8\pi x_{bj}^2 S_{ep}^2 Q^2}
\sum_{k=1}^4 {\cal A}_k \int_{x_{min}}^1\,{dx\over x}\int_{z_{min}}^1\,{dz\over z}\,
\nonumber\\
& &\qquad\qquad\times
\sum_{a}e_a^2 \left[ f_a(x) D_a(z)\widehat{\sigma}_k^{qq}
+ f_g(x) D_a(z)\widehat{\sigma}_k^{qg}+
f_a(x) D_g(z)\widehat{\sigma}_k^{gq}\right]\nonumber\\
& &\qquad\qquad\times\,\delta\left( {q_T^2\over Q^2} -
\left( {1\over \xhat} -1\right)\left({1\over \zhat}-1\right)\right) \; ,
\label{unpol}
\eeq
where $\alpha_{em}=e^2/4\pi$ is the QED coupling constant, 
$f_a(x)$ and $D_a(z)$ are, respectively, unpolarized quark distribution
and fragmentation functions with quark flavor $a$
and summation $\sum_a$ is over all quark and anti-quark flavors.
$f_g(x)$ and $D_g(z)$ are the gluon distribution and fragmentation functions, respectively. 
In (\ref{unpol}),   
we have introduced the variables
\beq
\xhat={x_{bj}\over x}\; ,\qquad \zhat={z_f\over z}\; ,
\eeq
and
\beq
x_{min}=x_{bj}\left( 1 + {z_f\over 1-z_f}{q_T^2\over Q^2}\right)\; ,\qquad
z_{min}=z_{f}\left( 1 + {x_{bj}\over 1-x_{bj}}{q_T^2\over Q^2}\right) \; .
\label{eq3.2}
\eeq
The partonic hard cross sections in (\ref{unpol}) are defined as ($C_F=(N_c^2 -1 )/(2N_c )$)
\beq
\widehat{\sigma}_1^{qq}&=&2C_F\xhat\zhat\left\{{1\over Q^2q_T^2}
\left({Q^4\over \xhat^2\zhat^2} + \left(Q^2-q_T^2\right)^2\right) +6\right\} \; ,
\nonumber\\
\widehat{\sigma}_2^{qq}&=&2\widehat{\sigma}_4^{qq}=8C_F\xhat\zhat \; ,\nonumber\\
\widehat{\sigma}_3^{qq}&=&4C_F\xhat\zhat{1\over Qq_T}(Q^2+q_T^2) \; ,
\label{eq3.avqq}
\eeq
\beq
\widehat{\sigma}_1^{qg}&=&\xhat(1-\xhat)\left\{{Q^2\over q_T^2}
\left({1\over \xhat^2\zhat^2} -{2\over \xhat\zhat}
+2\right) +10 -{2\over \xhat}-{2\over \zhat}\right\}\; ,
\nonumber\\
\widehat{\sigma}_2^{qg}&=&2\widehat{\sigma}_4^{qg}=8\xhat(1-\xhat) \; ,\nonumber\\
\widehat{\sigma}_3^{qg}&=&\xhat(1-\xhat){2\over Qq_T}\left\{2(Q^2+q_T^2)
-{Q^2\over \xhat\zhat}\right\} \; ,
\label{eq3.avgq}
\eeq
\beq
\widehat{\sigma}_1^{gq}&=&2C_F\xhat(1-\zhat)\left\{{1\over Q^2q_T^2}
\left({Q^4\over \xhat^2\zhat^2} + {(1-\zhat)^2\over \zhat^2}
\left(Q^2-{\zhat^2q_T^2\over (1-\zhat)^2}
\right)^2\right) +6\right\} \; ,
\nonumber\\
\widehat{\sigma}_2^{gq}&=&
2\widehat{\sigma}_4^{gq}=8C_F\xhat(1-\zhat) \; ,\nonumber\\
\widehat{\sigma}_3^{gq}&=&-4C_F\xhat(1-\zhat)^2{1\over \zhat Qq_T}
\left\{Q^2+
{\zhat^2 q_T^2\over (1-\zhat)^2}\right\} \; .
\label{eq3.avqg}
\eeq
For a given $S_{ep}$, $Q^2$ and $q_T$,
the kinematic constraints for $x_{bj}$ and $z_f$ are
\beq
& &{Q^2\over S_{ep}}<x_{bj}<1 \; ,
\label{range_zb}\\
& &0< z_f < {1-x_{bj}\over
1-x_{bj}+x_{bj}q_T^2/Q^2} \; .
\label{range_zf}
\eeq
Consequently, $q_T$ is limited by
\beq
0< q_T < Q\sqrt{\left({1\over x_{bj}}-1\right)\left({1\over z_{f}}-1\right)} \; .
\eeq
In a frame where $\vec{q}$ and $\vec{p}$ 
are collinear, the transverse momentum, $P_{hT}$, obeys
\beq
P_{hT}=z_f q_T <
z_fQ\sqrt{\left({1\over x_{bj}}-1\right)\left({1\over z_{f}}-1\right)} \; .
\label{zlimit}
\eeq

\section{Calculation of the single-spin-dependent cross section}
In Section~2, we have seen that
the single-spin-dependent cross section arises
from the pole part of the internal propagators in the partonic 
hard cross sections. 
When the 
conditions
(\ref{consistency1}) and (\ref{consistency2}) for the hard part
are satisfied in the Feynman gauge calculation,
the color gauge invariance of the cross section and the consistency of the calculational 
procedure are guaranteed.  
In this section, we shall show that these conditions are truly satisfied
by all types of pole contributions in the hard part, and derive the twist-3 factorization 
formula for the 
single-spin-dependent cross section.  
Because the relevant three types of poles arise at the different position in the phase 
space as in (\ref{HPprop})-(\ref{SGPprop}),
the 
conditions
(\ref{consistency1}) and (\ref{consistency2}) should be satisfied separately by the 
corresponding hard part
that arises from each pole contribution.
We first discuss in detail each pole contribution in the hard part for the quark 
fragmentation channel,
in particular, those associated with the quark-gluon correlation function $G_F$ for the nucleon 
in Section~3.1. The results for the gluon fragmentation channel as well as the 
contributions associated with
the correlation function $\GFt$ 
will be presented in the subsequent sections 3.2, 3.3.

\begin{figure}[t!]
\begin{center}
\epsfig{figure=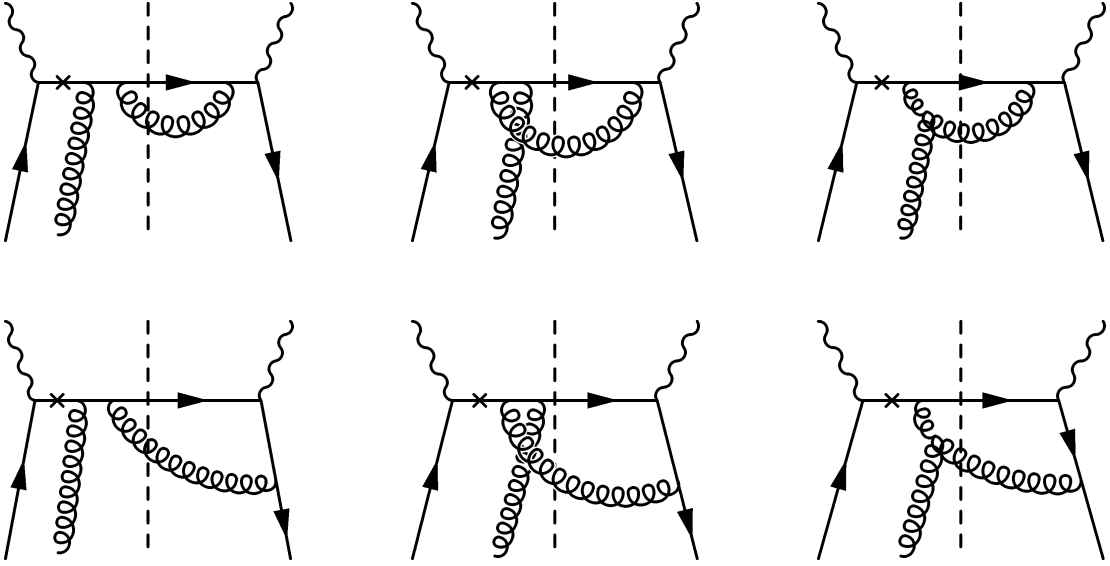,width=0.9\textwidth}
\end{center}
\caption{Feynman diagrams which give rise to the 
HP contributions in the quark fragmentation channel, where
the hard quark
fragments into final-state pion and the hard gluon goes into unobserved final state. 
The cross $\times$ denotes 
the quark propagator which gives the HP contribution.
Mirror diagrams also contribute.
\label{fig2}}
\end{figure}

\subsection{$G_F$ contribution: quark fragmentation channel}
\subsubsection*{A. Hard pole}
The diagrams for partonic subprocesses 
which produce hard-pole (HP) contributions are shown in Fig. 2,
with $k_{1,2}$ the momenta of ``external quarks'' as was defined in Fig. 1(b).  
The momentum flowing through the
quark line marked by a cross is $k_1+q$, and the corresponding quark propagator produces the HP
at $(k_1+q)^2 =0$, 
which reduces to the pole at $x_1=x_{bj}$ after the collinear expansion,
$k_1 \rightarrow x_1 p$ (see (\ref{HPprop})). 
The corresponding HP contribution can be evaluated as the imaginary part of the quark propagator 
through the distribution identity,
\begin{equation}
\frac{1}{ (k_1+q)^2 +i\varepsilon} = P\frac{1}{(k_1+q)^2} -i\pi \delta\left((k_1+q)^2 \right)\ ,
\label{ditrrr}
\end{equation}
before the collinear expansion.
In the following, we assume that the diagrams with a cross represent
the corresponding Feynman amplitude only with its pole contribution 
for the crossed propagator.

\begin{figure}[t!]
\begin{center}
\epsfig{figure=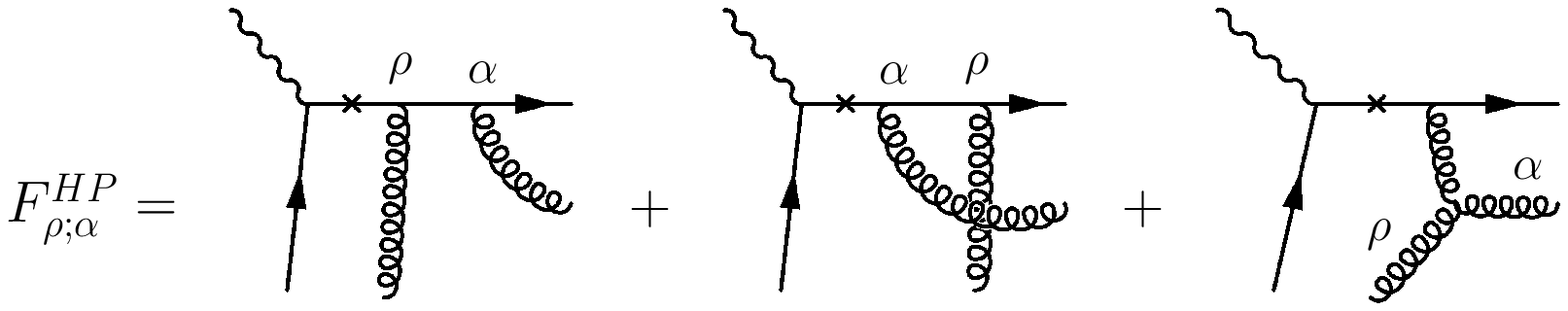,width=0.9\textwidth}
\end{center}
\caption{The definition of $F_{\rho;\alpha}^{HP}$, as the sum of three types of diagrams 
which appear in the LHS of the cut in Fig.~2.
\label{fig3}
}
\end{figure}

Fig.~3 shows the three types of different diagrams appearing in the LHS of the cut of the 
diagrams in Fig. 2, and we 
define $F^{HP}_{\rho;\alpha} (k_1,k_2,q,P_h/z)$ as the sum of 
the three diagrams in Fig.~3,
where the factors for the external lines are amputated,
$\rho$ is the 
Lorentz index for the coherent gluon from the nucleon, and $\alpha$ is the Lorentz index for 
the hard gluon emitted into the final state. 
$F^{HP}_{\rho;\alpha} (k_1,k_2,q,P_h/z)$ contains 
the factor $\delta\left((k_1+q)^2\right)$ from the second term of (\ref{ditrrr}).
Due to the tree level Ward identity shown in Fig. 4,
$F^{HP}_{\rho;\alpha} (k_1,k_2,q,P_h/z)$ satisfies the relation
\beq
(k_2-k_1)^\rho\ \bar{u}(P_h/z)F^{HP}_{\rho;\alpha} 
(k_1,k_2,q,P_h/z)\epsilon_{(\lambda)}^{*\alpha}(k_2+q-P_h/z)
\delta\left((k_2+q-P_h/z)^2\right)=0,
\label{HPWard}
\eeq
where $\bar{u}(P_h/z)$ is the spinor for the quark fragmenting into pion,
$\epsilon_{(\lambda)}^\alpha(k_2+q-P_h/z)$ is the polarization vector for the
hard gluon with the momentum $k_2+q-P_h/z$ and the physical polarization $\lambda$,
and $\delta\left((k_2+q-P_h/z)^2\right)$ comes from the on-shell condition
for the final gluon.  We also define the amputated amplitude
$f_\beta(k_2,q,P_h/z)$ for the sum of the diagrams shown in Fig. 5, which represent two types of 
diagrams 
in the RHS of the cut 
in Fig. 2.
\begin{figure}[t!]
\begin{center}
\epsfig{figure=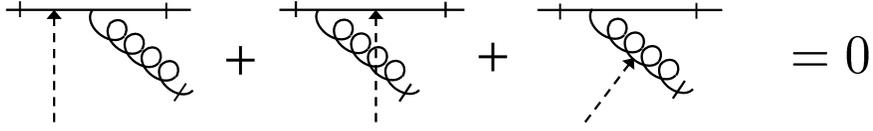,width=1.0\textwidth,clip=}
\end{center}
\caption{Ward identity for the HP contribution defined in Fig.~3.  
The dashed line represents a scalar-polarized gluon, and the parton lines marked by a bar
are on shell.
}
\end{figure}
\begin{figure}[t!]
\begin{center}
\epsfig{figure=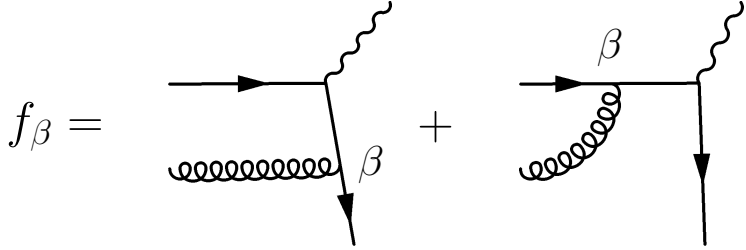,width=0.6\textwidth}
\end{center}
\caption{The definition of $f_\beta$, as the sum of two types of diagrams 
which appear in the RHS of the cut in Fig.~2.
}
\end{figure}
Combining $F^{HP}_{\rho;\alpha} (k_1,k_2,q,P_h/z)$ and $f_\beta(k_2,q,P_h/z)$, 
the sum of the six diagrams in Fig. 2 can be written as
\beq
\lefteqn{S^{HP}_\rho(k_1,k_2,q,P_h/z)}\nonumber\\
&&=
f_\beta(k_2,q,P_h/z)\,\frac{\sslash{P}_h}{z}\,F^{HP}_{\rho;\alpha} (k_1,k_2,q,P_h/z)\,
{\cal P}^{\alpha\beta}(k_2+q-P_h/z)\,
2\pi\,\delta\left((k_2+q-P_h/z)^2\right),\nonumber\\
\label{SHP}
\eeq
where we have used the relation
$\sum_{\rm spins}u(P_h/z) \bar{u}(P_h/z)= \sslash{P}_h /z$, and  
\beq
{\cal P}^{\alpha\beta}(k)\equiv\sum_\lambda
\epsilon_{(\lambda)}^{*\alpha}(k)\epsilon_{(\lambda)}^{\beta}(k)
\label{pola}
\eeq
is the polarization tensor for the on-shell ($k^2=0$) hard gluon
with the sum of $\lambda$ restricted over the physical polarizations.  
Note that (\ref{SHP}) is exactly the part of $S^{(1)}_{\alpha}(k_1,k_2,q,P_h/z)$
appearing in (\ref{w-twist3-f}), which comes from the HP contribution.
Owing to the relation (\ref{HPWard}), $S^{HP}_\rho(k_1,k_2,q,P_h/z)$ satisfies
\beq
(k_2-k_1)^\rho S^{HP}_\rho(k_1,k_2,q,P_h/z)=0.
\label{HPidentity}
\eeq
We now take the derivative of (\ref{HPidentity}) with respect to $k_i^\sigma$ ($i=1,2$;
$\sigma=\perp$) and set
$k_i=x_ip$.  One then obtains
\beq
& &(x_2-x_1) \left.{\partial S^{HP}_\rho(k_1,k_2,q,P_h/z)p^\rho
\over \partial k_2^\sigma}\right|_{k_i=x_ip} +S_\sigma^{HP}(x_1p,x_2p,q,P_h/z)=0,
\label{HPcond2}\\
& &(x_2-x_1)\left.{\partial S^{HP}_\rho(k_1,k_2,q,P_h/z)p^\rho
\over \partial k_1^\sigma}\right|_{k_i=x_ip}-S_\sigma^{HP}(x_1p,x_2p,q,P_h/z)=0.
\label{HPcond1}
\eeq
This proves that the relation (\ref{consistency1}) holds for the HP contribution.
Since $x_1\neq x_2$ for the HP contribution,
the relations (\ref{HPcond2}) and (\ref{HPcond1}) can be written as
\beq
& &\left.{\partial S^{HP}_\rho(k_1,k_2,q,P_h/z)p^\rho
\over \partial k_2^\sigma}\right|_{k_i=x_ip} = -
\left.{\partial S^{HP}_\rho(k_1,k_2,q,P_h/z)p^\rho
\over \partial k_1^\sigma}\right|_{k_i=x_ip} \nonumber\\
& & \qquad\qquad ={1\over x_1-x_2} S_\sigma^{HP}(x_1p,x_2p,q,P_h/z),
\label{HPcalc}
\eeq
which proves that the condition (\ref{consistency2}) is satisfied.

Furthermore,
the relation (\ref{HPcalc}) provides a useful method to calculate the corresponding hard 
cross section:   
Instead of calculating the derivative $\left.{\partial S^{HP}_\rho(k_1,k_2,q,P_h/z)p^\rho
\over \partial k_2^\sigma}\right|_{k_i=x_ip}$ 
for the first term of (\ref{w-twist3-f}), one simply needs to calculate 
$S_\sigma^{HP}(x_1p,x_2p,q,P_h/z)/(x_1-x_2)$, the RHS of (\ref{HPcalc}).
$S^{HP}_\rho(k_1,k_2,q, P_h/z)p^\rho$ represents the sum of the six diagrams
in Fig. 2, and, as was noticed in (\ref{HPWard}), (\ref{SHP})
above, the contribution from every diagram 
contains the product of two delta functions as
$\delta\left((k_1+q)^2\right)\delta\left((k_2+q-P_h/z)^2\right)$.
Accordingly the calculation of the derivative
$\left.{\partial S^{HP}_\rho(k_1,k_2,q,P_h/z)p^\rho
\over \partial k_2^\sigma}\right|_{k_i=x_ip}$
produces the derivative of these delta functions.
However, 
these terms with the derivative of delta functions eventually cancel among each other, since
$S_\sigma^{HP}(x_1p,x_2p,q,P_h/z)/(x_1-x_2)$ does not contain such derivatives.
This explains the absence of the derivatives of delta functions in the HP contribution,
which has been observed in the literature.~\footnote{The authors of \cite{JQVY06,JQVY06DIS} 
calculated 
$\left.{\partial S^{HP}_\rho(k_1,k_2,q,P_h/z)p^\rho
/ \partial k_2^\sigma}\right|_{k_i=x_ip}$ directly
to obtain the HP contributions for Drell-Yan and SIDIS, and found the cancellation
among the terms with the derivative of the delta functions.}
Therefore,
it is much more convenient to calculate
$S_\sigma^{HP}(x_1p,x_2p,q,P_h/z)/(x_1-x_2)$ to obtain the hard cross section 
for the 
HP contribution.

We also note that the calculation of
$S_\sigma^{HP}(x_1p,x_2p,q,P_h/z)$ can be further simplified.  
It is defined as (\ref{SHP}) in terms of the
physical polarization tensor (\ref{pola}) which can be expressed as
\beq
{\cal P}^{\alpha\beta}(k)
= -g^{\alpha\beta}+ { k^\alpha h^\beta + h^\alpha k^\beta
\over h\cdot k},
\label{Polarization}
\eeq
where $h^\alpha$ is an arbitrary light-like vector ($h^2=0$) 
satisfying $h\cdot k \neq 0$.  However, for $S_\sigma^{HP}(x_1p,x_2p,q,P_h/z)$ 
with the transverse Lorentz index $\sigma=\perp$, which guarantees
the physical polarization for the coherent gluon with the momentum $(x_2-x_1)p$ in the 
diagrams in Fig. 2, 
one can 
make the replacement 
${\cal P}^{\alpha\beta}(x_2p+q-P_h/z)
\to -g^{\alpha\beta}$ because the contribution of the second term of (\ref{Polarization}) vanishes 
by the Ward identities owing to the gauge invariance for the sum of the relevant diagrams. 

It is straightforward to check that electromagnetic gauge invariance, 
$q_\mu w^{\mu \nu} =q_\nu w^{\mu \nu} =0$ for
our hadronic tensor (\ref{w-twist3-f}), 
is satisfied by the HP contributions.
Substituting (\ref{HPcalc}) and (\ref{twist3distr}) into the first term of (\ref{w-twist3-f}), 
the relevant hard cross section associated with 
$G_F (x_1 , x_2 )$ is
proportional to ${\rm Tr}\left[S_\sigma^{HP}(x_1p,x_2p,q,P_h/z)\pslash\right]/(x_1-x_2)$.  
Using the Ward identities of QED,
the sum of the diagrams in Fig.~2 for 
${\rm Tr}\left[S_\sigma^{HP}(x_1p,x_2p,q,P_h/z)\pslash\right]$
vanishes, when a virtual photon vertex is contracted by $q_\mu$ or $q_{\nu}$.

By the method described above, it is now straightforward to calculate the
HP contribution to the cross section.

\begin{figure}[t!]
\begin{center}
\epsfig{figure=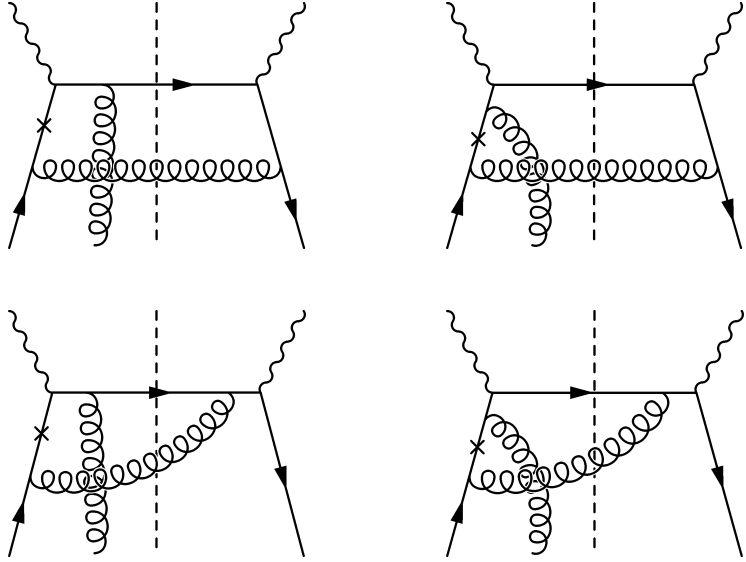,width=0.6\textwidth}
\end{center}
\caption{Same as Fig.~2, but for the SFP contributions in the quark fragmentation channel.
\label{fig6}
}
\end{figure}

\subsubsection*{B. Soft-fermion pole}

The diagrams which give rise to the soft-fermion-pole (SFP) contributions are
shown in Fig. 6, similarly to those for the HP in Fig.~2.
The internal quark propagator which produces the SFP is indicated by 
a cross in each diagram of Fig. 6. The momentum flowing through this line
is $k_1-k_2+P_h/z -q$, so that the SFP arises at $(k_1-k_2+P_h/z -q)^2 =0$.
Since each graph accompanies the
factor $\delta\left((k_2+q-P_h/z)^2\right)$ coming from the
cut propagator for the hard gluon,
this pole reduces to that at $x_1=0$ as in (\ref{SFPprop}) in the collinear limit 
$k_{1,2} \rightarrow x_{1,2} p$.
The SFP contribution can be evaluated as the imaginary part 
$\propto\delta\left((k_1-k_2+P_h/z -q)^2\right)$ of the corresponding propagator 
before the collinear
expansion.
As in the case of the Ward identity (\ref{HPWard}) for the HP contribution,
the relevant Ward identity for the present case is shown in Fig. 7. 
Following the same step as for the HP contribution,
it is easy to prove that the 
conditions
(\ref{consistency1}) and (\ref{consistency2}) are satisfied by the SFP contribution,
and also (\ref{HPcalc}) with $HP \rightarrow SFP$ holds.
Accordingly, the SFP contribution can be calculated via 
$S_\perp^{SFP}(x_1p,x_2p,q,P_h/z)/(x_1 -x_2)$ and 
${\cal P}^{\alpha\beta}
\to -g^{\alpha\beta}$, i.e., 
in the same economic way as in the HP contribution. 
It is also straightforward to check that electromagnetic gauge invariance is 
satisfied by the SFP contributions
using the Ward identities of QED, i.e.,
the sum of the diagrams in Fig.~6 
for the calculation of 
${\rm Tr}\left[ S_\perp^{SFP}(x_1p,x_2p,q,P_h/z)\pslash\right]/(x_1-x_2)$ vanishes
when a virtual photon vertex is contracted by $q_\mu$ or $q_{\nu}$.

\begin{figure}[t!]
\begin{center}
\epsfig{figure=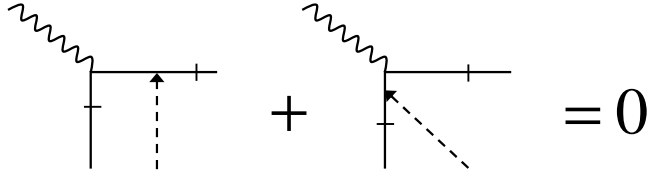,width=1.0\textwidth,clip=}
\end{center}
\caption{Ward identity for the SFP contribution, similarly to Fig.~4.
\label{fig7}
}
\end{figure}

The method discussed above gives the HP and SFP contributions
in terms of the $F$-type functions of (\ref{twist3distr}) based on 
(\ref{w-twist3-f})-(\ref{consistency2}).
It is instructive to note that the HP and SFP contributions can be rewritten
in terms of the $D$-type distributions of (\ref{twist3D}) using the relation (\ref{FDvector}). 
In the new expression, 
$S_\perp^{HP,SFP}(x_1p,x_2p,q,P_h/z)$ can be regarded as the hard part associated
with the $D$-type distributions.  

In this connection, we note that
the authors
of \cite{QS91}, in their study on $p^\uparrow p\to\gamma X$, 
employed a procedure of calculating $S_\perp^{SFP}(x_1 p,x_2p,q,P_h/z)$ directly
as the hard cross section corresponding to the $D$-type 
distributions, and eventually obtained the results which
are formally the same as those obtained by the above mentioned
rewriting of the $F$-type functions in terms of the $D$-type functions.
However, the logic for this procedure in \cite{QS91} is not based on the relations like
(\ref{HPcalc}), (\ref{FDvector}), (\ref{FDrelation}) in contrast to our method.
In fact, the authors of \cite{QS91} 
tried to find the hard cross sections corresponding to
the $D$-type and $F$-type distributions independently, and
relied on their claim
that the $F$-type functions do not receive the SFP contribution
(see Section 3.3 of \cite{QS91}).  
But this statement cannot be consistent with the relation (\ref{FDvector}) which was 
not noticed in  \cite{QS91}:
If this statement were true, it would mean that the $D$-type functions
do not receive the SFP contribution either, 
since the $F$-type and $D$-type functions are proportional to each other
as shown in (\ref{FDvector}) and (\ref{FDrelation}) for $x_1\neq x_2$.  
Their argument leading to the above claim was the following: The $A^+$ component 
of the coherent gluon field represents the scalar-polarized component 
at the relevant accuracy
in the collinear expansion, 
and thus 
decouples from a gauge-invariant set of diagrams in the hard part
by means of the Ward identity, so that the hard part 
for the $F$-type function, 
$\left. \{ {\partial S^{(1)}_\rho(k_1,k_2,q,P_h/z)p^\rho
/ \partial k_2^\perp}-{\partial S^{(1)}_\rho(k_1,k_2,q,P_h/z)p^\rho
/ \partial k_1^\perp} \} \right|_{k_i=x_ip}$, identified as the coefficient of the 
combination $\partial^{\perp} A^{+}$, 
should vanish at $x_1\neq x_2$ . (And 
the hard part for the $F$-type 
function can survive only for the SGP contributions with  $x_1=x_2$, 
where $A^+$ does not represent a scalar-polarized component.)
However, the correct relations from the Ward identity are those as given in 
(\ref{HPcond2}) and (\ref{HPcond1}), which are different from
their relation.  
Our demonstration leading to (\ref{HPcond2}) and (\ref{HPcond1})
indicates that it is not allowed to apply the Ward identities naively to the ``derivative
of the amplitude'' which appears as a next-to-leading order term 
in the collinear expansion. 
We emphasize that both ingredients ------ (i) the relations (\ref{FDvector}), (\ref{FDrelation}) 
among the nonperturbative functions,
and (ii) the correct relations (\ref{HPcond2}), (\ref{HPcond1}) from 
the Ward identities for the partonic subprocess ------ 
are essential to guarantee the consistency of the calculation. 
Using only one of them would lead to contradiction or
double counting.

\begin{figure}[t!]
\begin{center}
\epsfig{figure=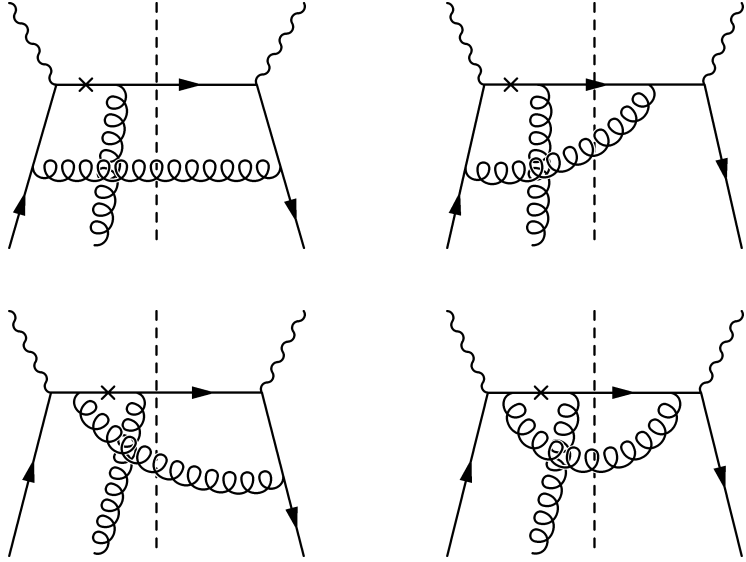,width=0.6\textwidth}
\end{center}
\caption{Same as Fig.~2, but for the SGP contributions in the quark fragmentation channel.
\label{fig8}
}
\end{figure}

\subsubsection*{C. Soft-gluon pole}

The diagrams which give rise to the soft-gluon-pole (SGP) contributions are shown in Fig. 8, 
where
the quark line which produces the pole is indicated by a cross.  
The momentum flowing through this line is $k_1-k_2+P_h/z$, so that the SGP arises
at $(k_1-k_2+P_h/z)^2 = 0$, which reduces to the pole at $x_1=x_2$ as in (\ref{SGPprop})
for $k_{1,2} \rightarrow x_{1,2}p$.
The SGP contribution can be evaluated as the imaginary part
$\propto\delta\left((k_1-k_2+P_h/z)^2\right)$ of the corresponding propagator, i.e., 
taking the 
quark line on-shell.

Define $S^{L}_{\rho}(k_1,k_2,q,P_h/z)$ as the sum of the diagrams shown in Fig. 8, where
the factors for the external lines are amputated, 
the index $\rho$ is for the coherent gluon, and the 
superscript $L$ indicates that
the coherent gluon is attached to the LHS of the cut.
Likewise we can also define $S^{R}_{\rho}(k_1,k_2,q,P_h/z)$ 
corresponding to the mirror diagrams of Fig.~8. 
$S^{L,R}_{\rho}(k_1,k_2,q,P_h/z)$ represent the SGP contributions in the corresponding 
partonic hard scattering
as directed to take the quark line on-shell by the crosses
in Fig. 8. It also has the delta function as the cut propagator for the on-shell hard gluon.
Therefore $S^{L}_{\rho}(k_1,k_2,q,P_h/z)$ is proportional to 
the product of the two delta functions as
\beq
S^{L}_{\rho}(k_1,k_2,q,P_h/z)\sim
\delta\left((k_1-k_2+P_h/z)^2\right)\delta\left((k_2+q-P_h/z)^2\right).
\label{SGPdelta}
\eeq
Notice that we are considering $S^{L}_{\rho}(k_1,k_2,q,P_h/z)$ with
parton momenta {\it before} collinear expansion, as in the HP and SFP cases discussed above.   
Since the quark-gluon vertex attaching the coherent gluon is sandwiched by
the on-shell quark lines, the tree level Ward identity implies
\beq
(k_2-k_1)^\rho S^{L}_{\rho}(k_1,k_2,q,P_h/z)=0.
\label{SGPWard}
\eeq
Actually, the Ward identity of this type holds for each diagram 
in Fig. 8. 
By taking the derivative of (\ref{SGPWard}) with respect to
$k_i^\sigma$ ($i=1,2$; $\sigma =\perp$) and setting $k_i=x_ip$, one arrives at
\beq
& &(x_2-x_1) \left.{\partial S^{L}_\rho(k_1,k_2,q,P_h/z)p^\rho
\over \partial k_2^\sigma}\right|_{k_i=x_ip} +S_\sigma^{L}(x_1p,x_2p,q,P_h/z)=0,
\label{SGPcond2}\\
& &(x_2-x_1)\left.{\partial S^{L}_\rho(k_1,k_2,q,P_h/z)p^\rho
\over \partial k_1^\sigma}\right|_{k_i=x_ip}-S_\sigma^{L}(x_1p,x_2p,q,P_h/z)=0.
\label{SGPcond1}
\eeq
Apparently the similar relations hold for $S^{R}_\rho(k_1,k_2,q,P_h/z)$, too.
Noting that $S^{L}_\rho(k_1,k_2,q,P_h/z)$
$+S^{R}_\rho(k_1,k_2,q,P_h/z)$
is the part of $S^{(1)}_\rho(k_1,k_2,q,P_h/z)$ appearing in (\ref{w-twist3-f}), which comes from
the SGP contribution,
the above
equation (\ref{SGPcond2}) shows that the required consistency condition
(\ref{consistency1}) is satisfied.

So far, the logic
used for deriving (\ref{SGPcond2}) and (\ref{SGPcond1})
is the same as that used for the HP and SFP cases. 
However, there is an important difference between (\ref{SGPcond2}), (\ref{SGPcond1})
and (\ref{HPcond2}), (\ref{HPcond1}).
Namely, we cannot derive the relation similar to (\ref{HPcalc}) 
using (\ref{SGPcond2}) and (\ref{SGPcond1}), because the first term of (\ref{SGPcond2}), 
(\ref{SGPcond1})
contains singularity at $x_1 = x_2$ due to the delta function. 
Owing to (\ref{SGPdelta}), 
$\left.{\partial S^{L}_\rho(k_1,k_2,q,P_h/z)p^\rho
/ \partial k_2^\sigma}\right|_{k_i=x_ip}$ consists of the 
terms with 
$\delta'\left(x_1-x_2\right)\delta\left((x_2p+q-P_h/z)^2\right)$, 
$\delta\left(x_1-x_2\right)\delta'\left((x_2p+q-P_h/z)^2\right)$ or
$\delta\left(x_1-x_2\right)\delta\left((x_2p+q-P_h/z)^2\right)$. 
Because of the 
factor $x_2-x_1$
in the first term 
of (\ref{SGPcond2}), 
the calculation of 
$S_\sigma^{L}(x_1p,x_2p,q,P_h/z)$ misses the terms
in $\left.{\partial S^{L}_\rho(k_1,k_2,q,P_h/z)p^\rho
/ \partial k_2^\sigma}\right|_{k_i=x_ip}$, which
vanish due to $(x_1-x_2)\delta(x_1-x_2)=0$.\,\footnote{
This also explains why it is difficult to obtain the SGP cross section 
in the lightcone gauge, $A^+=0$.  In this gauge, one would
identify the contribution of the $F$-type functions
as $A^\perp\to iF^{\perp n}/(x_1-x_2)$ and calculate
$S_\perp^{L,R}(x_1p,x_2p,q,P_h/z)$ to get the cross section.  
But a naive calculation of  
$S_\perp^{L,R}(x_1p,x_2p,q,P_h/z)$ would miss some parts of the SGP
cross section as was noted above.}
Therefore 
it is necessary to
calculate $\left.{\partial S^{L}_\rho(k_1,k_2,q,P_h/z)p^\rho
/ \partial k_2^\sigma}\right|_{k_i=x_ip}$ directly
to get the SGP contribution to the cross section correctly.

\begin{figure}[t!]
\begin{center}
\epsfig{figure=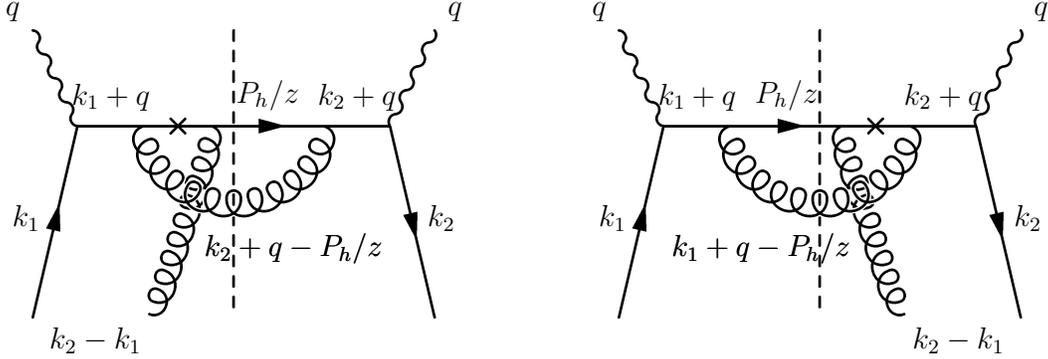,width=0.85\textwidth}
\end{center}
\caption{An example of a pair of diagrams from Fig.~8, 
with the coherent gluon attached on the different side of the cut.
}
\end{figure}

Because of the lacking of the analogue of (\ref{HPcalc}),
we have to rely on direct inspection of the diagrams in Fig. 8 and their mirror diagrams 
in order to show that the condition (\ref{consistency2}) is realized 
for the SGP contribution.  
We first note that, 
when the derivative $\partial/\partial k_{1,2}^\alpha$
hits an internal propagator which depends on $k_{1}$ and $k_2$ 
only through $k_1-k_2$, the condition (\ref{consistency2}) is automatically satisfied
for that contribution. 
For other cases, one needs to
consider a combination of some diagrams in Fig. 8 and their mirror diagrams.  
As an example, consider a pair of the diagrams shown in Fig.~9.  
The momentum flowing through the quark line with a cross are, respectively, 
$k_1-k_2+P_h/z$ and $k_2-k_1+P_h/z$ for Fig. 9 (a) and (b), and they give SGPs which have 
opposite signs of the residue in the collinear limit, $k_i\rightarrow x_ip$ ($i=1,2$). 
On the other hand, Fig. 9 (a) and (b), respectively, contain 
$\delta\left((k_2+q-P_h/z)^2\right)$ and
$\delta\left((k_1+q-P_h/z)^2\right)$ as the cut propagator for the on-shell hard gluon. 
Therefore, for the sum of these two diagrams in Fig.~9, 
the derivative hitting on these delta functions satisfies 
(\ref{consistency2}) at the SGP.
If the derivative hits other internal propagator such as
that with momentum $k_1+q$, or $k_2+q$, in Fig. 9, the corresponding contribution simply 
cancel between the diagrams, because
the SGPs have opposite signs of the residue between the diagrams 
(a) and (b) in Fig. 9.
It is straightforward to see that the action of the derivative $\partial/\partial k_{1,2}^\alpha$
can be classified 
into one of these three cases for all diagrams in Fig.~8.
Therefore, the condition (\ref{consistency2}) is also satisfied
for the SGP contribution by taking into account all the graphs
including the mirror diagrams.  
This is in contrast to the HP and SFP cases, where
the condition (\ref{consistency2}) is already satisfied
by summing the diagrams which have a coherent-gluon attachment
on one side of the cut.

Here we mention 
the form of the polarization tensor from the polarization sum (\ref{pola}) of 
the final-state hard gluon
in $S^{L,R}_\rho(k_1,k_2,q,P_h/z)$.  In principle, ${\cal P}^{\alpha\beta}(k_2+q-P_h/z)$
with (\ref{Polarization})
should be used 
to calculate $S^{L}_\rho(k_1,k_2,q,P_h/z)$, and   
likewise for $S^{R}_\rho(k_1,k_2,q,P_h/z)$. 
However, one is allowed to make the replacement 
${\cal P}^{\alpha\beta}(k_2+q-P_h/z) \rightarrow -g^{\alpha\beta}$.
This is formally similar to the HP and SFP cases, but the logic to allow the replacement 
is somewhat
sophisticated 
compared with those cases because we have to treat the scalar polarized coherent-gluon
in the present case.
For the proof, 
we first note that, 
in calculating the hard cross section associated with the quark-gluon correlation function $G_F$ 
based on
(\ref{w-twist3-f}), 
the corresponding spinor trace in the first term of (\ref{w-twist3-f})
is taken with the insertion of
the Dirac structure $\pslash$ of the first term of (\ref{twist3distr}).
We consider the 
following replacement for this $\pslash$ in the calculation: 
\beq
\pslash={1\over 2}\pslash\nslash\pslash\to {1\over 2x_1x_2}\kslash_1\nslash\kslash_2. 
\label{replace}
\eeq
A difference between the original hard cross section,
\beq
\left.{\partial\over\partial k_2^\sigma}
{\rm Tr}\left[S^{L,R}_\rho(k_1,k_2,q,P_h/z)p^\rho\pslash\right]
\right|_{k_i=x_ip}\ ,
\label{Soriginal}
\eeq
and the new one with (\ref{replace}),
\beq
\left.{\partial\over\partial k_2^\sigma}
{\rm Tr}\left[S^{L,R}_\rho(k_1,k_2,q,P_h/z)p^\rho{1\over 2x_1x_2}\kslash_1\nslash
\kslash_2\right]
\right|_{k_i=x_ip}\ ,
\label{Sreplaced}
\eeq
occurs when the derivative $\partial /\partial k_2^\sigma$
hits the $\kslash_{2}$ in (\ref{replace}).
However, these terms cancel between $S^L_\rho p^\rho$ and $S^R_\rho p^\rho$
for (\ref{Sreplaced}),
because the SGP appears with opposite signs of the residue between them.
Thus one can obtain the hard cross section using (\ref{Sreplaced}).  
We note that, 
in (\ref{Sreplaced}),
$\kslash_1$ ($\kslash_2$) 
plays the role of the factor for 
the incoming (outgoing) external quark line with the momentum
$k_1$ ($k_2$).  
It is now easy to see that,
with the on-shell condition $k_{1,2}^2=0$,\footnote{ 
In general, $k_{i}^2 = 2k_{i}^{+}k_{i}^{-}-{\mathbf{k}}_{i \perp}^2 \neq 0$ ($i=1,2$). 
Up to the twist-3 accuracy, however, we can choose as
$k_{i}^{-} = {\mathbf{k}}_{i \perp}^2 /(2k_{i}^{+})$, so that
our external quarks are taken on-shell $k_{i}^2 = 0$ here.}
the terms proportional to
$(k_{2}+q-P_h/z)^{\alpha}$ or $(k_{2}+q-P_h/z)^{\beta}$ in ${\cal P}^{\alpha\beta}(k_2+q-P_h/z)$
drops in
${\rm Tr}\left[S^{L,R}_\rho(k_1,k_2,q,P_h/z)p^\rho{1\over 2x_1x_2}\kslash_1\nslash
\kslash_2\right]$
with the aid of the tree level Ward identity.  
So one can make the replacement ${\cal P}^{\alpha\beta}(k_2+q-P_h/z) \to -g^{\alpha\beta}$
in calculating (\ref{Sreplaced}).  
Since there is no change 
between (\ref{Soriginal}) and (\ref{Sreplaced}) even
with this $-g^{\alpha\beta}$, 
one can use (\ref{Soriginal}) with the polarization tensor 
$-g^{\alpha\beta}$ for the final-state hard gluon.  
In this way, one can safely use $-g^{\alpha\beta}$ instead of (\ref{Polarization})
for the calculation of the SGP contribution based on
(\ref{w-twist3-f}).

Eq.~(\ref{Sreplaced}), combined with the on-shell condition $k_{1,2}^2=0$,
is also useful to prove the electromagnetic gauge invariance,
$q_\mu w^{\mu \nu} =q_\nu w^{\mu \nu} =0$, for the SGP contribution.
Using the Ward identities of QED,
the sum of the diagrams in Fig.~8 for the calculation of (\ref{Sreplaced}) 
vanishes when a virtual photon vertex is contracted by $q_\mu$ or $q_{\nu}$: 
The sum of vertical two diagrams for 
(\ref{Sreplaced}) 
vanishes when the left virtual photon vertex is contracted with $q_\mu$,
because of the on-shell incoming quark line with the factor $\kslash_1$ and the on-shell 
internal quark line with the SGP.
And the sum of the horizontal two diagrams vanishes when the right virtual photon 
vertex is contracted with $q_\mu$, because of the on-shell outgoing quark line with the factor $\kslash_2$ and 
the on-shell quark line emerging from the fragmentation insertion with $\Pslash_h$.

\subsubsection*{D. Result for the factorization formula}
Following the method described above, 
the single-spin-dependent cross section
associated with the twist-3 distribution for the nucleon, $G_F(x,y)$, 
and the quark fragmentation function for the pion can be obtained as
\beq
& &\frac{d^5\sigma^{Vq}}{dx_{bj}dQ^2dz_fdq_T^2d\phi}=
\frac{\alpha_{em}^2\alpha_S}{8\pi x_{bj}^2S_{ep}^2Q^2}
\sum_{k=1}^4 {\cal A}_k\left(\frac{-\pi M_N}{4}\right)\sin\Phi_S
\int_{x_{min}}^1\frac{dx}{x}\int_{z_{min}}^1\frac{dz}{z}
\nonumber \\
&&\qquad \quad \times \sum_a e_a^2\delta_a\left[
\hat{\sigma}^{Vq}_{Dk} \left(x \frac{d}{dx}G_F^a(x,x)\right)
+\hat{\sigma}^{Vq}_{Gk}\, G_F^a(x,x)\right. \nonumber \\
&&\qquad \qquad +\,\hat{\sigma}^{Vq}_{Hk}\, G_F^a(x_{bj},x)
+\hat{\sigma}^{Vq}_{Fk}\, G_F^a(0,x)\biggr]\,D_a(z)\,
\delta\left(\frac{q_T^2}{Q^2}-\left(1-\frac{1}{\hat{x}}\right)\left(
1-\frac{1}{\hat{z}}\right)\right),
\label{rrr1}
\eeq
where $G_F^a$ represents 
$G_F$ for quark-flavor $a$ with 
electric charge $e_a$, $\delta_a$ is defined as
$\delta_a=1$ for quark and $\delta_a=-1$ for anti-quark, and the summation $\sum_a$
is over all quark and anti-quarks.  
The factor ``$-\pi M_N /4$'' arises as the product of the factor $i$ in the first term of (\ref{w-twist3-f}),
$M_N /4$ in (\ref{twist3distr}), and $i\pi$ associated with evaluation of the pole contributions.
$\hat{\sigma}^{Vq}_{Dk}$, $\hat{\sigma}^{Vq}_{Gk}$, 
$\hat{\sigma}^{Vq}_{Hk}$ and $\hat{\sigma}^{Vq}_{Fk}$ are
the hard cross sections corresponding, respectively, to the derivative term for the 
SGP contribution, non-derivative term for the SGP contribution, HP contribution, and
SFP contributions.  They are given as
\beq
&&\hat{\sigma}^{Vq}_{Dk}=-\frac{4C_q q_T}{Q^2}
\frac{\hat{x}}{1-\hat{z}}\hat{\sigma}^{Vq}_{Uk},\nonumber \\ 
&&\qquad \qquad \left\{\begin{array}{l}
\hat{\sigma}^{Vq}_{U1} =\displaystyle{2 \hat{x}\hat{z} 
\left[ \frac{1}{Q^2 q_T^2} \left(\frac{Q^4}{\hat{x}^2 \hat{z}^2}
	+(Q^2-q_T^2)^2\right)+6 \right]},\\ \\
	\hat{\sigma}^{Vq}_{U2}=2 \hat{\sigma}^{Vq}_{U4}=8 \hat{x}\hat{z},\\ \\
	\hat{\sigma}^{Vq}_{U3}=\displaystyle{4 \hat{x}\hat{z}\frac{1}{Q q_T}(Q^2+q_T^2)},
	\end{array}\right. 
\label{qqSGPD}\\[8pt]
&&\left\{\begin{array}{l}
\hat{\sigma}^{Vq}_{G1}=\displaystyle{\frac{8C_q q_T}{Q^2}\hat{x} \left[
	\frac{1+\hat{z}^2}{(1-\hat{x})^2 (1-\hat{z})^2}
	+\frac{2 \hat{z}(2-3 \hat{z})+(1-2 \hat{x})(1+6 \hat{z}^2-6 
\hat{z})}{(1-\hat{z})^2}\right]}, \\ \\
\hat{\sigma}^{Vq}_{G2}=\displaystyle{2 \hat{\sigma}^{Vq}_{G4}=
\frac{64C_q q_T}{Q^2}\frac{\hat{x}^2 \hat{z}}{1-\hat{z}}}, \\ \\
\hat{\sigma}^{Vq}_{G3}=\displaystyle{\frac{8C_q}{Q}\hat{x}\left[
\frac{\hat{x}\hat{z}(5-4\hat{x})}{(1-\hat{x})(1-\hat{z})}+3-4 \hat{x}\right]},
\end{array}\right. 
\label{qqSGPS}\\[8pt]
&&\left\{\begin{array}{l}
\hat{\sigma}^{Vq}_{H1}=\displaystyle{\frac{8q_T}{Q^2}\left(C_F \hat{z} +{1\over 2N_c} \right)
\frac{\hat{x}(1+\hat{x}\hat{z}^2)}{(1-\hat{x})^2(1-\hat{z})^2}}, \\ \\
\hat{\sigma}^{Vq}_{H2}=0, \\ \\
\hat{\sigma}^{Vq}_{H3}=\displaystyle{\frac{8}{Q}
\left(C_F \hat{z} +{1\over 2N_c} \right)\frac{\hat{x}\hat{z}}{(1-\hat{x})(1-\hat{z})}}, \\ \\
\hat{\sigma}^{Vq}_{H4}=\displaystyle{\frac{8q_T}{Q^2}
\left(C_F \hat{z} +{1\over 2N_c} \right)\frac{\hat{x}\hat{z}}{(1-\hat{x})(1-\hat{z})}},
\end{array}\right. \\[8pt]
&&\left\{\begin{array}{l}
\hat{\sigma}^{Vq}_{F1}=\displaystyle{-\frac{8C_qq_T}{Q^2}\left[
-\frac{8 \hat{x}^2 \hat{z}}{1-\hat{z}}+\hat{x}(2 \hat{x}-1)
+\frac{\hat{x}^2 \hat{z}^2(2 \hat{x}^2-3 \hat{x}+1)}{(1-\hat{x})^2(1-\hat{z})^2}\right]}, \\ \\
\hat{\sigma}^{Vq}_{F2}=\displaystyle{\frac{64C_qq_T}{Q^2}\frac{\hat{x}^2 \hat{z}}{1-\hat{z}}}, \\ \\
\hat{\sigma}^{Vq}_{F3}=\displaystyle{-\frac{8C_q}{Q}\left[
\hat{x}(4 \hat{x}-3)-\frac{\hat{x}\hat{z}(4 \hat{x}-1)}{1-\hat{z}}\right]}, \\ \\
\hat{\sigma}^{Vq}_{F4}=\displaystyle{-\frac{8C_qq_T}{Q^2}\frac{\hat{x}\hat{z}
(1-2 \hat{x})^2}{(1-\hat{x})(1-\hat{z})}},
\end{array}\right. 
\label{qqSFP}
\eeq
where $C_q=-1/2N_c$ and 
$\hat{\sigma}^{Vq}_{Uk}$ in (\ref{qqSGPD})
is the same as (\ref{eq3.avqq}) for the unpolarized cross section (\ref{unpol})
except for the color factor.  
$\hat{\sigma}^{Vq}_{Dk}$ was already given in \cite{EKT06}.  
The above result for $\hat{\sigma}^{Vq}_{G1}$ and $\hat{\sigma}^{Vq}_{H1}$ 
also agree with those in \cite{JQVY06DIS}.
All the other results in (\ref{qqSGPS})-(\ref{qqSFP}) are new.

\begin{figure}[t!]
\begin{center}
\epsfig{figure=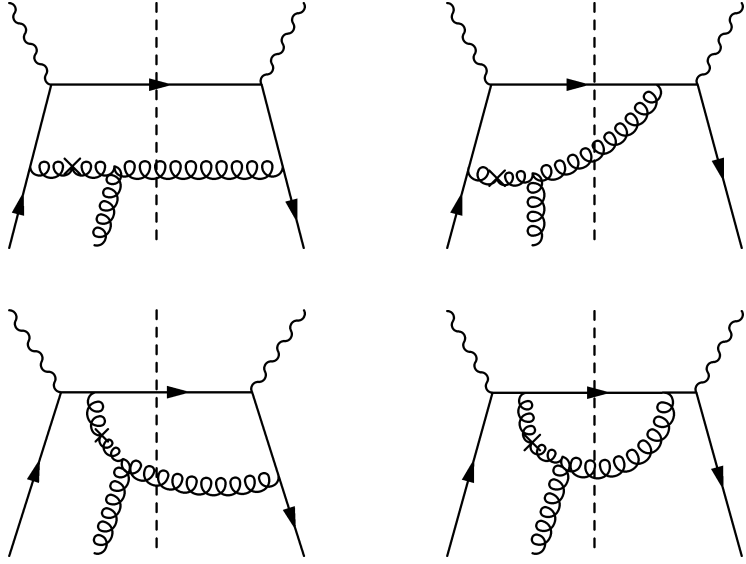,width=0.6\textwidth}
\end{center}
\caption{Same as Fig.~2, but for the SGP contributions in the gluon fragmentation channel
where the hard gluon 
fragments into final-state pion and the hard quark goes into unobserved final state. 
}
\end{figure}

\subsection{$G_F$ contribution: gluon fragmentation channel}
The twist-3 distribution $G_F(x,y)$ contributes to the single-spin-dependent cross section also
with the gluon fragmentation function for the pion, $D_g(z)$.
Again, this is 
classified into the HP, SFP and SGP contributions.
The diagrams corresponding to the HP and SFP contributions 
are given by the same diagrams as shown Figs.~2 and 6, respectively, except that 
the hard gluon 
fragments into final-state pion and the hard quark goes into unobserved final state. 
The diagrams for the SGP contributions 
are shown in Fig.~10.  
The method of the calculation and the proof of 
the conditions
(\ref{consistency1}) and (\ref{consistency2}) are the same as that
described in the previous Section~3.1.
So we shall not repeat it here.
The result is obtained as follows:  
\begin{eqnarray}
&&\frac{d^5\sigma^{Vg}}{dx_{bj}dQ^2dz_fdq_T^2d\phi}=
\frac{\alpha_{em}^2\alpha_S}{8\pi x_{bj}^2S_{ep}^2Q^2}
			\sum_{k=1}^4 {\cal A}_k\left(\frac{-\pi M_N}{4}\right)
\sin\Phi_S\int_{x_{min}}^1\frac{dx}{x}\int_{z_{min}}^1\frac{dz}{z}
\nonumber \\
&&\qquad \qquad\quad \times \sum_a e_a^2 \left[
\hat{\sigma}^{Vg}_{Dk}\, \left(x \frac{d}{dx}G_F^a(x,x)\right)
+\hat{\sigma}^{Vg}_{Gk}\, G_F^a(x,x)\right. \nonumber \\
&&\qquad\qquad\qquad 
+\delta_a\hat{\sigma}^{Vg}_{Hk}\, G_F^a(x_{bj},x)
		+\delta_a\hat{\sigma}^{Vg}_{Fk}\, G_F^a(0,x)\biggr]D_g(z)\,
\delta\left(\frac{q_T^2}{Q^2}-
\left(1-\frac{1}{\hat{x}}\right)\left(1-\frac{1}{\hat{z}}\right)
\right),\nonumber\\
\label{rrr2}
\eeq
where each hard cross section is given by
\beq
&&\hat{\sigma}^{Vg}_{Dk}=-\frac{4C_g q_T}{Q^2}
\frac{\hat{x}}{1-\hat{z}}\hat{\sigma}^{Vg}_{Uk},\nonumber \\ 
&&\qquad \qquad \left\{\begin{array}{l}
\hat{\sigma}^{Vg}_{U1} =\displaystyle{2\xhat(1-\zhat)\left\{{1\over Q^2q_T^2}
\left({Q^4\over \xhat^2\zhat^2} + {(1-\zhat)^2\over \zhat^2}
\left(Q^2-{\zhat^2q_T^2\over (1-\zhat)^2}
\right)^2\right) +6\right\}} ,\\ \\
	\hat{\sigma}^{Vg}_{U2}=2 \hat{\sigma}^{Vg}_{U4}= 8\xhat(1-\zhat) ,\\ \\
	\hat{\sigma}^{Vg}_{U3}=\displaystyle{-4\xhat(1-\zhat)^2{1\over \zhat Qq_T}
\left\{Q^2+
{\zhat^2 q_T^2\over (1-\zhat)^2}\right\}},
	\end{array}\right. 
\label{qgSGPD}\\[8pt]
&&\left\{\begin{array}{l}
\hat{\sigma}^{Vg}_{G1}=
\displaystyle{-C_g\frac{8 q_T}{Q^2}\frac{\hat{x}}{(1-\hat{x})^2(1-\hat{z})\hat{z}}}\\ \\
\qquad \qquad \times \displaystyle{\left[2(6 \hat{z}^2-6 \hat{z}+1)\hat{x}^3+(-24 \hat{z}^2+22 
\hat{z}-3)
\hat{x}^2+4 \hat{z}(3 \hat{z}-2)\hat{x}-\hat{z}^2-1 \right]},\\ \\
\hat{\sigma}^{Vg}_{G2}=
\displaystyle{2 \hat{\sigma}^{Vg}_{G4}=C_g\frac{64q_T}{Q^2}\hat{x}^2}, \\ \\
\hat{\sigma}^{Vg}_{G3}=\displaystyle{-C_g\frac{8}{Q}\frac{\hat{x}}{(1-\hat{x})\hat{z}}
\left[(8 \hat{z}-4)\hat{x}^2+(5-12 \hat{z})\hat{x}+3 \hat{z}\right]},
\end{array}\right.  \\[10pt]
&&\left\{\begin{array}{l}
\hat{\sigma}^{Vg}_{H1}=\displaystyle{\frac{8q_T}{Q^2}\left(C_F(\hat{z}-1) -{1\over 2N_c}\right)
			\frac{\hat{x}(1+\hat{x}(1-\hat{z})^2)}{(1-\hat{x})^2(1-\hat{z})\hat{z}}},\\ \\
\hat{\sigma}^{Vg}_{H2}=0, \\ \\
\hat{\sigma}^{Vg}_{H3}=\displaystyle{\frac{8}{Q}\left(C_F(\hat{z}-1)-{1\over 2N_c}\right)
			\frac{\hat{x}(\hat{z}-1)}{(1-\hat{x})\hat{z}}}, \\ \\
\hat{\sigma}^{Vg}_{H4}=\displaystyle{\frac{8q_T}{Q^2}\left(C_F(\hat{z}-1)-{1\over 2N_c}\right)
\frac{\hat{x}}{1-\hat{x}}},
\end{array}\right.\\[10pt]
&&\left\{\begin{array}{l}
\hat{\sigma}^{Vg}_{F1}=\displaystyle{-\frac{8C_qq_T}{Q^2}
\frac{\hat{x}}{(1-\hat{x})(1-\hat{z})\hat{z}}
			\left[2(6 \hat{z}^2-6 \hat{z}+1)\hat{x}^2+
(-12 \hat{z}^2+10 \hat{z}-1)\hat{x}+\hat{z}^2 \right]}, \\ \\
\hat{\sigma}^{Vg}_{F2}=\displaystyle{-\frac{64C_qq_T}{Q^2}\hat{x}^2}, \\ \\
\hat{\sigma}^{Vg}_{F3}=\displaystyle{-\frac{8C_q}{Q}\frac{\hat{x}}{\hat{z}}
			\left[-4 \hat{z}+\hat{x}(8 \hat{x}-4)+1 \right]}, \\ \\
\hat{\sigma}^{Vg}_{F4}=\displaystyle{\frac{8C_qq_T}{Q^2}\frac{(1-2 \hat{x})^2 
\hat{x}}{1-\hat{x}}},
\end{array}\right.
\eeq
where $C_g=N_c/2$.  
We remind that the hard cross section for the derivative term, (\ref{qgSGPD}),
was also derived in \cite{EKT06} and 
$\hat{\sigma}^{Vg}_{Uk}$ 
is the same as (\ref{eq3.avqg}) for the unpolarized cross section (\ref{unpol})
except for the color factor.  

\subsection{The $\GFt$ contribution}

Another twist-3 distribution for the nucleon, $\GFt (x,y)$, also contributes to 
$ep^\uparrow\to e\pi X$.  Again, in principle, the corresponding cross section arises as the 
HP, SFP and SGP
contributions.  
The diagrams for those contributions are the same as in the $G_F$ case
for the quark fragmentation channel as well as the gluon fragmentation channel, 
see Figs.~2, 6, 8, and 10. 
The calculation can be carried out in parallel with the case for $G_F$,
by changing the relevant Dirac matrix structure   
to calculate the first term of (\ref{w-twist3-f}) 
as $\pslash\epsilon^{\alpha pn S_\perp}\to
i \gamma_5\pslash S_\perp^\alpha$ (see (\ref{twist3distr})). 
Owing to the anti-symmetry (\ref{symF}),
one has $\GFt(x,x)=0$, so that there is no non-derivative term for the SGP contribution.
In principle, $\left.{\partial \GFt(x_1,x)\over \partial x_1}\right|_{x_1=x}$
can be nonzero and could contribute to
the cross section as the derivative term for the SGP contribution. 
However, after summing all the diagrams for the SGP contributions,
it turns out that no such derivative term survives. Therefore, there appears no SGP contribution, and 
the result for the single-spin-dependent cross section
associated with $\GFt (x, y)$ 
is obtained
from the HP and SFP contributions as 
\begin{eqnarray}
\frac{d^5\sigma^{A}}{dx_{bj}dQ^2dz_fdq_T^2d\phi}&=&\frac{\alpha_{em}^2\alpha_S}
{8\pi x_{bj}^2S_{ep}^2Q^2}
\sum_{k=1}^4 {\cal A}_k
\left(\frac{-\pi M_N}{4}\right)\sin\Phi_S\int_{x_{min}}^1\frac{dx}{x}\int_{z_{min}}^1\frac{dz}{z}
\nonumber \\
&&
\times\sum_a e_a^2\delta_a\left[
\left\{
\hat{\sigma}^{Aq}_{Hk}\, \widetilde{G}_F^a(x_{bj},x)
+\hat{\sigma}^{Aq}_{Fk}\, \widetilde{G}_F^a(0,x)\right\}\,D_a(z)\right.\nonumber\\
& &\left.+\left\{
\hat{\sigma}^{Ag}_{Hk}\, \widetilde{G}_F^a(x_{bj},x)
+\hat{\sigma}^{Ag}_{Fk}\, \widetilde{G}_F^a(0,x)\right\}\,D_g(z)\right]\nonumber\\
& &\times
\delta\left(\frac{q_T^2}{Q^2}-\left(1-\frac{1}{\hat{x}}\right)\left(1-\frac{1}{\hat{z}}\right)
\right),
\label{rrr3}
\eeq
where the hard cross sections for the quark fragmentation contribution are given by
\beq
&&\left\{\begin{array}{l}
\hat{\sigma}^{Aq}_{H1}=\displaystyle{\frac{8q_T}{Q^2}\left(C_F \hat{z} +{1\over 2N_c} \right)
\frac{\hat{x}(1-\hat{x}\hat{z}^2)}{(1-\hat{x})^2(1-\hat{z})^2}}, \\ \\
\hat{\sigma}^{Aq}_{H2}=0, \\ \\
\hat{\sigma}^{Aq}_{H3}=-\displaystyle{\frac{8}{Q}\left(C_F \hat{z} +{1\over 2N_c} \right)
\frac{\hat{x}\hat{z}}{(1-\hat{x})(1-\hat{z})}}, \\ \\
\hat{\sigma}^{Aq}_{H4}=-\displaystyle{\frac{8q_T}{Q^2}\left(C_F \hat{z} +{1\over 2N_c} \right)
\frac{\hat{x}\hat{z}}{(1-\hat{x})(1-\hat{z})}},
\end{array}\right. \\[5pt]
&&\left\{\begin{array}{l}
\hat{\sigma}^{Aq}_{F1}=\displaystyle{\frac{8C_qq_T}{Q^2}\frac{\hat{x}(1-\hat{z})^2+
\hat{x}(2 \hat{z}-1)}{(1-\hat{x})(1-\hat{z})^2}}, \\ \\
\hat{\sigma}^{Aq}_{F2}=0, \\ \\
\hat{\sigma}^{Aq}_{F3}=\displaystyle{\frac{8C_q}{Q}\frac{\hat{x}}{1-\hat{z}}}, \\ \\
\hat{\sigma}^{Aq}_{F4}=\displaystyle{\frac{8C_qq_T}{Q^2}
\frac{\hat{x}\hat{z}}{(1-\hat{x})(1-\hat{z})}},
\end{array}\right. 
\end{eqnarray}
and those for the gluon fragmentation contribution are given by
\beq
&&\left\{\begin{array}{l}
\hat{\sigma}^{Ag}_{H1}=\displaystyle{\frac{8q_T}{Q^2}
\left(C_F(\hat{z}-1) -{1\over 2N_c} \right)
\frac{\hat{x}(1-\hat{x}(1-\hat{z})^2)}{(1-\hat{x})^2(1-\hat{z})\hat{z}}}, \\ \\
\hat{\sigma}^{Ag}_{H2}=0, \\ \\
\hat{\sigma}^{Ag}_{H3}=\displaystyle{\frac{8}{Q}
\left(C_F(\hat{z}-1)-{1\over 2N_c} \right)\frac{\hat{x}(1-\hat{z})}{(1-\hat{x})\hat{z}}}, \\ \\
\hat{\sigma}^{Ag}_{H4}=\displaystyle{-\frac{8q_T}{Q^2}
\left(C_F(\hat{z}-1)-{1\over 2N_c} \right)\frac{\hat{x}}{1-\hat{x}}},
\end{array}\right.\\[5pt]
&&\left\{\begin{array}{l}
\hat{\sigma}^{Ag}_{F1}=\displaystyle{\frac{8C_qq_T}{Q^2}\frac{(-\hat{z}^2+2 
\hat{x}\hat{z}-\hat{x})\hat{x}}{(1-\hat{x})(1-\hat{z})\hat{z}}}, \\ \\
\hat{\sigma}^{Ag}_{F2}=0, \\ \\
\hat{\sigma}^{Ag}_{F3}=\displaystyle{\frac{8C_q}{Q}\frac{\hat{x}}{\hat{z}}}, \\ \\
\hat{\sigma}^{Ag}_{F4}=\displaystyle{-\frac{8C_qq_T}{Q^2}\frac{\hat{x}}{1-\hat{x}}}.
\end{array}\right.
\end{eqnarray}

\section{Azimuthal asymmetry}

By summing all the pole contributions with $G_F$ and $\GFt$ derived as (\ref{rrr1}), (\ref{rrr2}) and
(\ref{rrr3}) in Section~3,
one obtains the complete single-spin-dependent cross section for $ep^\uparrow\to e\pi X$
from the (A) term in (\ref{twist3}).
Its $\phi$-dependence can be decomposed as
\beq
{d^5\sigma^{({\rm A})}\over dx_{bj} dQ^2 dz_f dq_T^2 d\phi}
=\sin\Phi_S\left( \sigma_0^{\rm A} 
+\sigma_1^{\rm A} \cos(\phi) +\sigma_2^{\rm A}\cos(2\phi)\right).  
\label{azimuthA}
\eeq
This should be compared with the similar decomposition of the twist-2 unpolarized
cross section of (\ref{unpol})\,\cite{Mendez78,MOS92}: 
\beq
{d^5\sigma^{\rm unpol}\over dx_{bj} dQ^2 dz_f dq_T^2 d\phi}
=\sigma_0^U +\sigma_1^U \cos(\phi) +\sigma_2^U\cos(2\phi).  
\label{azimuthunpol}
\eeq
If one uses the {\it lepton plane} as a reference plane to
define the azimuthal angle of the spin vector
of the initial proton as $\phi_S$, and that of the hadron plane as $\phi_h$,
as employed in \cite{hermes}, one has the relation
$\Phi_S=\phi_S-\phi_h$ and $\phi=-\phi_h$.\footnote{In \cite{EKT06}, we made
a sign mistake in this relation.}
Accordingly, the azimuthal dependence of 
the $\sigma_0^{\rm A}$ term in (\ref{azimuthA}) is the same as the Sivers
effect ($\propto\sin(\phi_h-\phi_S)$) and (\ref{azimuthA}) contains 
the azimuthal components which are absent in the
leading order Sivers effect.  

We now present a simple estimate of SSA 
by using the obtained formulae. 
So far we don't have any definite information on the 
HP, SFP and SGP components of $G_F^a$ and $\GFt^a$.
In \cite{EKT06}, we presented an estimate
of SSA including only the derivative term of the SGP function, ${d\over dx}G_F^a(x,x)$,
which is expected to be a good approximation at
large $x_{bj}$ or large $z_f$. 
For comparison with that study, we will here include all the
SGP contribution, the derivative and non-derivative terms of the SGP function.
This calculation may be relevant, 
since we expect that there are much more soft gluons in the nucleon than 
the gluons with finite momentum.  

In order to see the relative importance of the
non-derivative term, we assume the same model
for $G^a_F(x,x)$ as in \cite{EKT06}:  
We take into account only the contributions from $u$ and $d$ quarks
with the ansatz $G_F^a(x,x)=K_a f_a(x)$ ($a=u,d$)\,\cite{Koike03} 
and the flavor-dependent constant
$K_u=-K_d=0.07$, which were shown to approximately reproduce 
the observed SSA of $p^\uparrow p\to \pi X$ \cite{E704,STAR}
as SGP contributions. 
As in \cite{EKT06}, 
we calculate the azimuthal asymmetries normalized by the unpolarized cross section.
We define $\phi$-integrated azimuthal asymmetries at $\Phi_S=\pi/2$ as, 
\beq
\la 1 \ra_N \equiv {\sigma_0^{\rm A} \over \sigma_0^U},\qquad
\la\cos\phi\ra_N\equiv {\sigma_1^{\rm A}\over 2\sigma_0^U},\qquad 
\la\cos 2\phi\ra_N\equiv {\sigma^{\rm A}_2\over 2\sigma_0^U}.  
\label{ssaA}
\eeq
For the purpose of illustration, we choose two sets of 
kinematic variables: 
The first one is $S_{ep}=300$ GeV$^2$, $Q^2=20$ GeV$^2$, $x_{bj}=0.1$, 
which is close to the COMPASS kinematics.  Another one is
$S_{ep}=2500$ GeV$^2$, $Q^2=50$ GeV$^2$ and $x_{bj}=0.03$, which is in the region of
planned eRHIC experiment\,\cite{erhic}.  
Both sets give the same $\cosh\psi$ in (\ref{eq2.cosh}).  
In our calculation,
we have used GRV unpolarized parton distribution $f_{a,g}(x)$\,\cite{GRV98}
and KKP pion fragmentation function $D_{a,g}(z)$\,\cite{KKP00}
as in \cite{EKT06}.

\begin{figure}[t!]
\begin{center}
\epsfig{figure=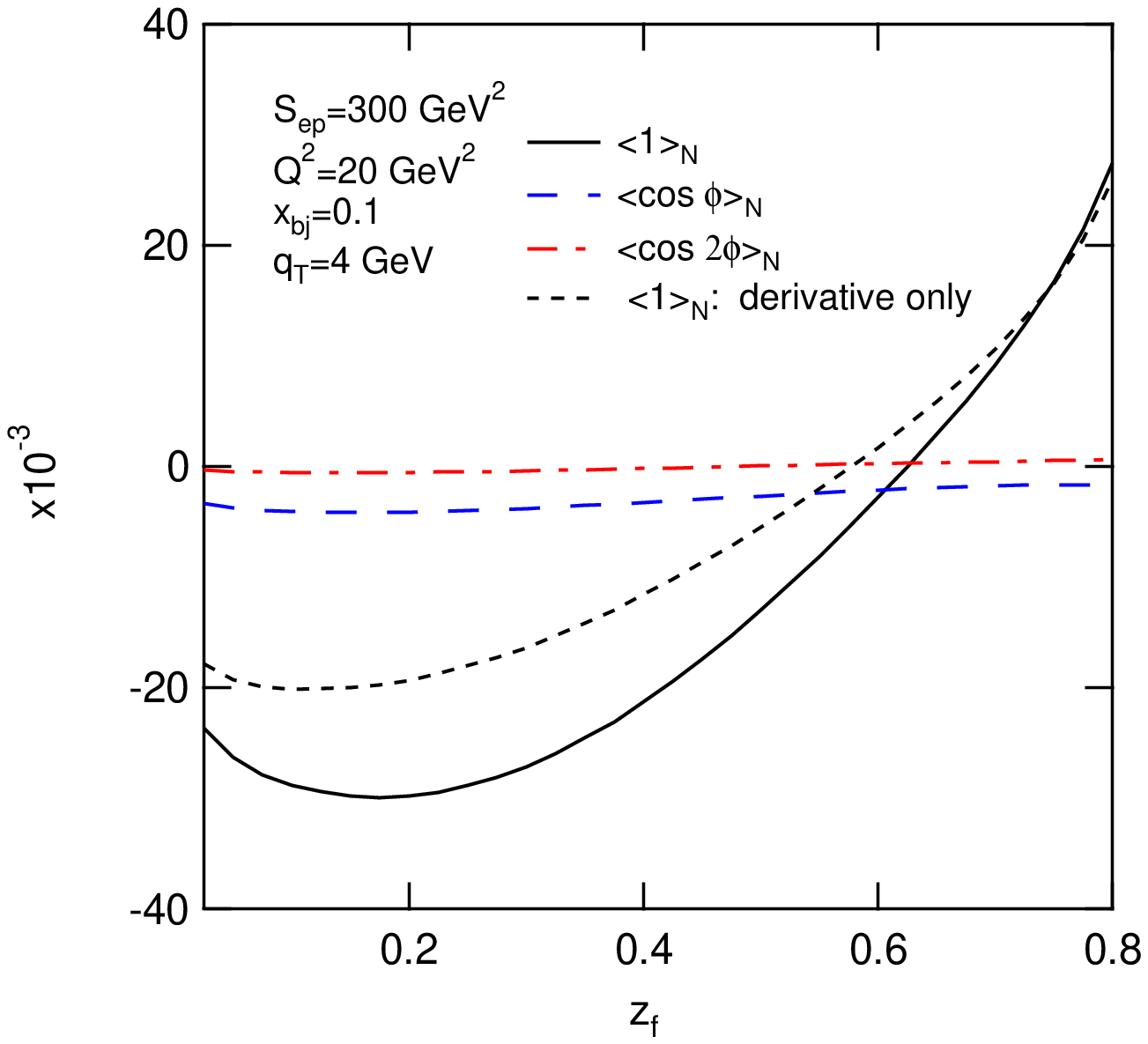,width=0.48\textwidth}
\epsfig{figure=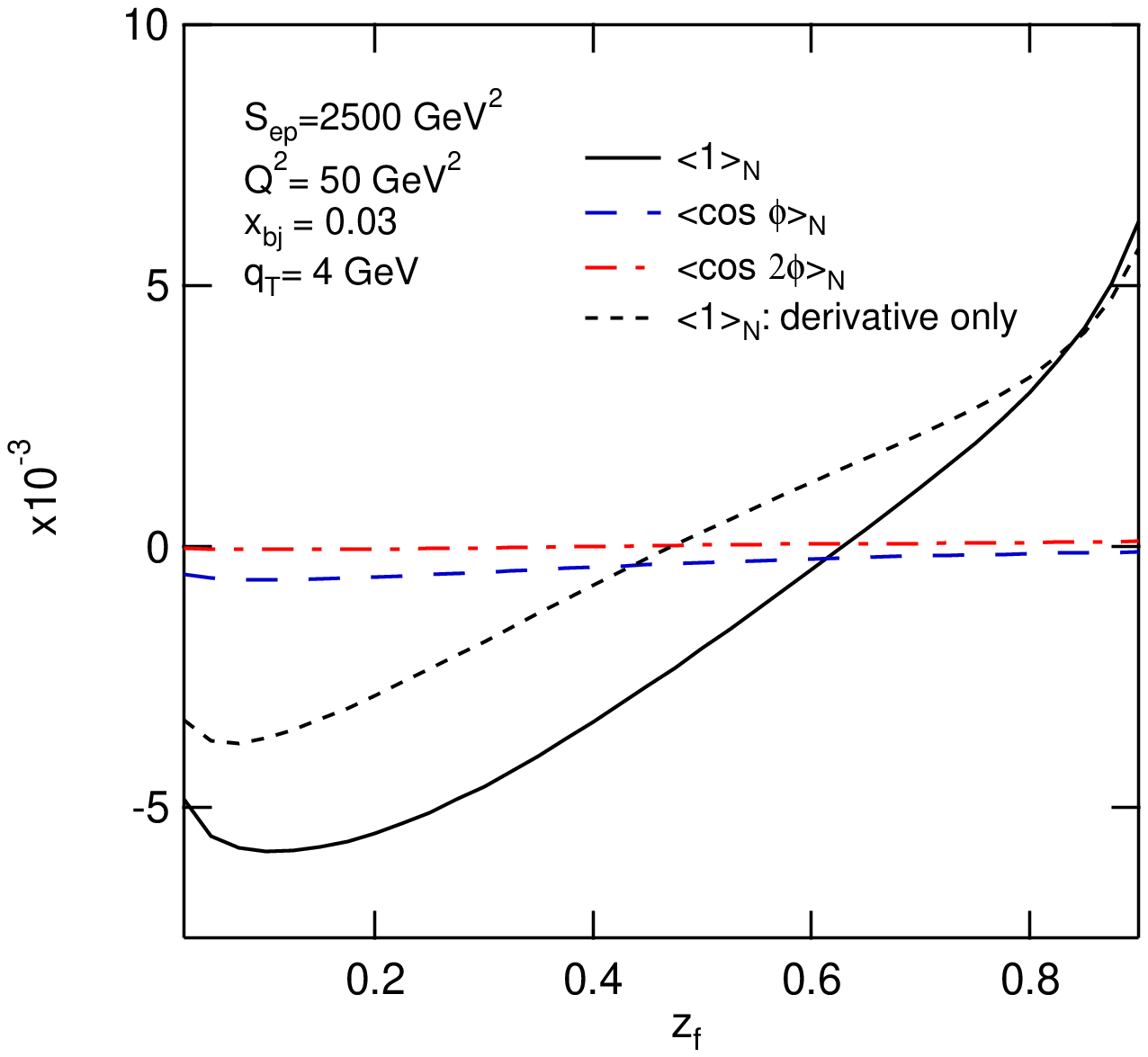,width=0.48\textwidth}
\end{center}
\vspace{-0.5cm}
\hspace{4cm}(a)
\hspace{7.5cm}(b)
\caption{
The calculated azimuthal asymmetries (\ref{ssaA}) with 
all the SGP contributions for
(a) the COMPASS kinematics and (b) the eRHIC kinematics. 
For $\la 1\ra_N$, we have also shown the result with only the derivative term of the 
SGP contribution. 
}
\end{figure}

The results for the
azimuthal asymmetries (\ref{ssaA}) for the $\pi^0$ production in SIDIS are
shown in Figs. 11(a) and (b) for the COMPASS and the eRHIC kinematics,
respectively. The asymmetries $\la 1\ra_N$, $\la\cos\phi\ra_N$, 
and $\la\cos2\phi\ra_N$ are shown by the solid, long-dashed, 
and dot-dashed curves, respectively,
at $q_T=4$ GeV as a function of $z_f$.
For comparison, 
we also showed, by the short-dashed curve, the result for $\la 1 \ra_N$ which was obtained only 
with the derivative term of the SGP contribution. 
One sees that $\la 1\ra_N$ is the order of 2 \%
for the COMPASS kinematics and is the order of 0.5 \% for the eRHIC kinematics, 
while $\la\cos\phi\ra_N$ and  $\la\cos2\phi\ra_N$ are negligible for
both kinematics.  
One also sees, comparing the solid and short-dashed curves, that the effect of the non-derivative terms of the
SGP contribution is important in the small and intermediate $z_f$ region, while the
derivative term dominates in the large $z_f$ region.
In the large $z_f$ region, where the large $x$ region of the distribution function becomes important (see (\ref{eq3.2}),
(\ref{rrr1}), (\ref{rrr2})),
the derivative term $(d/dx)G_F^a(x,x)$ together with 
our ansatz $G_F^a(x,x)=K_a f_a(x)$ causes a steep rising behavior of the asymmetry.  
If one had used a model for $G_F^a(x,x)$ which behaves as $\sim (1-x)^{\beta}$ at $x\to 1$
with larger $\beta$ than the present model $G_F^a(x,x)=K_a f_a(x)$, 
as suggested by a recent study on $p^\uparrow p\to \pi X$ \cite{KQVY06},
one would not have got such a strongly rising behavior and would have obtained a negative 
asymmetry for $\la 1 \ra_N$ in the whole $z_f$ region.  
It is quite natural to expect that the negative trend of the asymmetry
$\la 1\ra_N$ persists in the lower energy region, which is consistent with the 
recent HERMES data\,\cite{hermes}. (Recall the difference between the present work and \cite{hermes}
in the sign convention of the relevant angles 
for the asymmetry, as stated just below (\ref{azimuthunpol}).)

It has been known that the effect of the soft gluon radiation is important
in the shown $q_T$ region.  However, recent study on these effects 
for the twist-2 double spin asymmetry based on the
$q_T$ resummation formalism shows that the resummation effect more or less cancels
between the polarized and unpolarized cross sections and the fixed order 
perturbative result for the asymmetry gives a reasonable prediction
for the asymmetry\,\cite{KNV06}.
Although the resummation formalism for the twist-3 cross section has not yet been 
developed, it should be the one which resums logarithmically enhanced contributions
due to the emission of the soft gluons, whose coupling is generically blind to spin degrees of freedom.
Accordingly, we expect the result shown in Fig. 11 provides a correct order of magnitude
for the SSA (\ref{ssaA}).

\section{Summary and conclusion}

In this paper, we have studied the SSA for the pion 
production in SIDIS, $ep^\uparrow\to e\pi X$,
in the framework of the collinear factorization.
This study is relevant for the pion production
with large transverse momentum.  
In this framework, the SSA is a twist-3 observable, and 
the single-spin-dependent cross section occurs as pole contributions 
from internal propagators in the partonic subprocess, which produce the strong interaction phase necessary for SSA.  
We have derived the complete cross section formula associated with
the twist-3 quark-gluon correlation functions 
for the transversely polarized nucleon, by
including all types of pole (hard-pole, soft-fermion-pole and 
soft-gluon-pole) contributions. 
To accomplish this study,
we have proved that the hard part from each pole contribution satisfies 
certain constraints from the Ward identities for color gauge invariance, 
and also clarified a complete set of the gauge-invariant, twist-3 correlation functions.
These new developments are crucial
to prove the factorization property for the twist-3 single-spin-dependent cross section
in manifestly gauge-invariant form,
and the ``uniqueness'' of the corresponding factorization formula, which
was missing in the literature.  
Based on the soft-gluon-pole approximation, we have presented 
a simple estimate for the azimuthal asymmetry of SSA. 
It turned out that
the non-derivative term of the SGP contribution causes sizable
correction to the results using the derivative term only, for the small as well as moderate values of 
$x_{bj}$ and $z_f$.

Our formalism of the twist-3 calculation 
can be extended
to study SSAs in other processes of current interest such as 
$p^\uparrow p\to\gamma X$, 
$p^\uparrow p\to\pi X$, 
which will be reported in the future publication.

\section*{Acknowledgements}
We are grateful to J. Qiu and W. Vogelsang for useful discussions. 
The work of K.T. was 
supported by the Grant-in-Aid for Scientific Research No. C-16540266.


\end{document}